\documentstyle[preprint,eqsecnum,aps,pra,amssymb,amsfonts,newlfont,amsmath,tracefnt]{revtex}

\begin{document}
\draft

\title{Post-Newtonian approximation for isolated systems\\
calculated by matched asymptotic expansions}

\author{Olivier Poujade and Luc Blanchet}
\address{Institut d'Astrophysique de Paris (C.N.R.S.),\\ 
98\textsuperscript{~\!bis} boulevard Arago, 75014 Paris, France}

\date{\today}

\maketitle

\begin{abstract}
Two long-standing problems with the post-Newtonian approximation for
isolated slowly-moving systems in general relativity are~: (i) the
appearance at high post-Newtonian orders of divergent Poisson
integrals, casting a doubt on the soundness of the post-Newtonian
series; (ii) the domain of validity of the approximation which is
limited to the near-zone of the source, and prevents one, {\it a
priori}, from incorporating the condition of no-incoming radiation, to
be imposed at past null infinity. In this article, we resolve the
problem (i) by iterating the post-Newtonian hierarchy of equations by
means of a new (Poisson-type) integral operator that is free of
divergencies, and the problem (ii) by matching the post-Newtonian
near-zone field to the exterior field of the source, known from
previous work as a multipolar-post-Minkowskian expansion satisfying
the relevant boundary conditions at infinity. As a result, we obtain
an algorithm for iterating the post-Newtonian series up to any order,
and we determine the terms, present in the post-Newtonian field, that
are associated with the gravitational-radiation reaction onto an
isolated slowly-moving matter system.
\end{abstract}



\newtheorem{theorem}{Theorem}
\newtheorem{lemma}{Lemma}
\newtheorem{proposition}{Proposition}
\newtheorem{definition}{Definition}

\newcommand{\Dt}{\widetilde{\Delta^{-1}}}
\newcommand{\Dn}[1]{\widetilde{\Delta^{#1}}}
\newcommand{\Dlt}{\widetilde{\Box^{-1}_\mathrm{Ret}}}
\newcommand{\vc}[1]{{\mathbf{\mathrm{#1}}}}
\newcommand{\Real}{\mathbb{R}}
\newcommand{\Z}{\mathbb{Z}}
\newcommand{\N}{\mathbb{N}}
\newcommand{\C}{\mathbb{C}}
\newcommand{\un}{\underset}
\newcommand{\pf}{\un{B=0}{\mathrm{FP}}~\!}
\newcommand{\qp}{\frac{1}{4\pi}}
\newcommand{\mn}{^{\mu\nu}}
\newcommand{\pn}[1]{\overline{#1}}
\newcommand{\pmk}{{\cal M}}
\newcommand{\xl}[1]{\hat{#1}_L}
\newcommand{\dl}[1]{\hat{\partial}_L({#1})}
\newcommand{\im}{{\cal I}^{-1}}
\newcommand{\dx}{d^3{\mathbf x}}
\newcommand{\dy}{d^3{\mathbf y}}
\newcommand{\dz}{dz}
\newcommand{\rmn}{r^{-1}}
\newcommand{\absx}{|{\mathbf x}|}
\newcommand{\absy}{|{\mathbf y}|}
\newcommand{\abty}{|{\widetilde{\mathbf y}}|}


\section{Introduction}\label{I}

\subsection{Problems with the post-Newtonian expansion}\label{IA}

The post-Newtonian approximation, or expansion when the speed of light
$c\to +\infty$, has been formalized in the early days of general
relativity by Einstein \cite{einstein}, Droste \cite{droste}, and
DeSitter \cite{desitter}.  Since then, it has provided us with our
best insights about the problems of motion and gravitational
radiation, two of general relativity's most important
issues. Concerning the problem of motion, we quote the dynamics of $N$
separated bodies at the first post-Newtonian (1PN, or $1/c^2$) order~:
works of Einstein, Infeld and Hoffmann \cite{EIH} and other authors
\cite{LD17,Fock39,Fock}, and the dynamics of extended fluid systems up
to the 2.5PN level of gravitational radiation reaction~: works of
Chandrasekhar and collaborators \cite{C65,CN69,CE70} and followers
\cite{AD75,Ehl77,Ehl80,Ker80,Ker80',Capo81,PapaL81,BRu81,BRu82}.  In
the case of two compact objects, we know the 2.5PN equations of motion
of the binary pulsar \cite{DD81a,D82,Dhouches,BFP98}, and the 3PN
equations of motion of inspiralling compact binaries
\cite{JaraS98,DJS00,BF00,BFeom,ABF01}. The specific contribution of
the gravitational-radiation reaction has been obtained up to the 1.5PN
relative order by the method of matched asymptotic expansions for
extended fluids \cite{BuTh70,Bu71,BD88,B93,B97}, and by means of
balance equations for compact binary systems
\cite{IW93,IW95}. Concerning the problem of gravitational radiation,
the work has focused on the expressions of the multipole moments of
general fluid systems \cite{BD86,BD89,DI91a,B95,B98mult}, and on the
gravitational-wave flux emitted by inspiralling compact binaries,
including the specific effects of wave tails, up to the 3.5PN order
\cite{BDI95,WWi96,B96,B98tail,BIJ01}.

The ``standard'' post-Newtonian approximation, at the basis of most of
the body of work quoted previously, is known to be plagued with some
apparently inherent difficulties, which crop up at some high
post-Newtonian order like 3PN.  Up to the 2.5PN order the
approximation can be worked out without problems, and at the 3PN order
the problems can be solved specifically for each case at hands (see
for instance Ref. \cite{BFeom}).  However, it must be admitted that
these difficulties, even appearing at higher approximations, cast
doubt on the actual soundness, on the theoretical point of view, of
the post-Newtonian expansion.  What is maybe worse, they pose the
practical question of the reliability of this approximation when
comparing the theory's predictions with very precise experimental
results. It is therefore highly desirable to assess the nature of
these difficulties -- are they purely technical or linked with some
fundamental drawback of the approximation scheme? -- and eventually to
resolve them.  This is especially important in view of the fact that
inspiralling compact binaries, when they are detected and analyzed by
gravitational-wave experiments, will necessitate a prior theoretical
knowledge of the gravitational-wave signal at some very high
post-Newtonian order \cite{BDI95,WWi96,B96,B98tail,BIJ01}.  In this
article let us distinguish (and resolve) the two basic problems faced
by the post-Newtonian expansion.

The first problem is that in higher approximations some {\it
divergent} Poisson-type integrals appear.  Recall that the
post-Newtonian expansion replaces the resolution of an hyperbolic-like
d'Alembertian equation by a perturbatively equivalent hierarchy of
elliptic-like Poisson equations.  Rapidly it is found during the
post-Newtonian iteration that the right-hand-side of the Poisson
equations acquires a non-compact support (it is distributed over all
space), and that the standard Poisson integral diverges because of the
bound of the integral at spatial infinity, i.e. $r\equiv |{\mathbf x}|\to
+\infty$, with $t=$ const.  For instance some of the potentials
occuring at the 2PN order in Chandrasekhar's work \cite{CN69} are
divergent, so the corresponding metric is formally infinite \footnote{Nevertheless, these divergencies were not a problem when
considering the equations of motion because the gradients of these
potentials, which parametrize the equations, were finite.}
. In fact, Kerlick \cite{Ker80,Ker80'} showed that the
post-Newtonian computation {\it \`a la} Chandrasekhar
\cite{C65,CN69,CE70}, following the iteration scheme of Anderson and
DeCanio \cite{AD75}, can be made well-defined up to the 2.5PN order,
by keeping some derivatives inside some crucial integrals to make them
finite \cite{Ehl77,Ehl80}.  However, the latter remedy does not solve
the problem at the next 3PN order, which has been found to involve
some inexorably divergent Poisson integrals \cite{Ker80,Ker80'}.

These divergencies come from the fact that the post-Newtonian
expansion is actually a singular perturbation, in the sense that the
coefficients of the successive powers of $1/c$ are not uniformly valid
in space, since they typically blow up at spatial infinity like some
positive powers of $r$. For instance, Rendall \cite{Rend92} has shown
that the post-Newtonian expansion cannot be ``asymptotically flat''
starting at the 2PN or 3PN level, depending on the adopted coordinate
system. The result is that the Poisson integrals are in general
badly-behaving at infinity. Physically this can be understood by the
fact that the post-Newtonian approximation is valid only in the near
zone of the source (see below) while the Poisson integral extends over
the whole three-dimensional space, including the regions far from the
source where the approximation breaks down. Therefore, trying to solve
the post-Newtonian equations by means of the standard Poisson integral
does not {\it a priori} make sense. This does not mean that there are
no solution to the problem, but simply that the Poisson integral does
not constitute the correct solution of the Poisson equation in the
context of post-Newtonian expansions.  So the difficulty is purely of
a technical nature, and will be solved once we succeed in finding the
appropriate solution to the Poisson equation \footnote{The problem is somewhat similar to what happens in
Newtonian cosmology. Here we have to solve the Poisson equation
$\Delta U=-4\pi G\rho$, where the density $\rho$ of the cosmological
fluid is constant all over space~: $\rho=\rho(t)$.  Clearly the
Poisson integral of a constant density does not make sense, as it
diverges at the bound at infinity like the integral $\int r dr$. This
nonsensical result has occasionally been referred to as the ``paradox
of Seeliger''.  However the problem is solved once we realize that the
Poisson integral does not constitute the appropriate solution of the
Poisson equation in the context of Newtonian cosmology. A well-defined
solution is simply given by $U=-\frac{2}{3}\pi G\rho r^2$.}. A solution to the problem of divergencies has been proposed by Futamase and Schutz \cite{FS83} and Futamase \cite{F83}. Their approach is alternative to the one we shall follow below. It is based on an initial-value formalism, which avoids the appearance of divergencies because of the finiteness of the integration region.

The second problem has to do with the near-zone limitation of the
approximation. Indeed the post-Newtonian expansion assumes that all
retardations $r/c$ are small, so it can be viewed as a formal {\it
near-zone} expansion when $r\to 0$, which is valid only in the region
surrounding the source that is of small extent with respect to the
typical wavelength of the emitted radiation~: $r\ll \lambda$ (if we
locate the origin of the coordinates $r=0$ inside the
source). Therefore, the fact that the coefficients of the
post-Newtonian expansion blow up at spatial infinity, when $r\to
+\infty$, has nothing to do with the actual behaviour of the field at
infinity.  The serious consequence is that it is not possible, {\it a
priori}, to implement within the post-Newtonian iteration the physical
information that the matter system is isolated from the rest of the
universe. Most importantly, the no-incoming radiation condition,
imposed at past null infinity, cannot be taken into account, {\it a
priori}, into the scheme.  In a sense the post-Newtonian approximation
is not ``self-supporting'', because it necessitates some information
taken from outside its own domain of validity.

To the lowest post-Newtonian orders one can circumvent this difficulty
by considering {\it retarded} integrals that are formally expanded
when $c\to +\infty$ as series of ``instantaneous'' Poisson-like
integrals \cite{AD75}. This procedure works well up to the 2.5PN level
and has been shown to correctly fix the dominant radiation reaction
term at the 2.5PN order \cite{Ker80,Ker80'}. Unfortunately such a
procedure assumes fundamentally that the gravitational field, after
expansion of all retardations $r/c\to 0$, depends on the state of the
source at a single time, in keeping with the instantaneous character
of the Newtonian interaction. However, we know that from the 4PN order
the post-Newtonian field (as well as the source's dynamics) ceases to
be given by a functional of the source parameters at a single time,
because of the imprint of gravitational-wave tails in the near zone
field, in the form of some modification, at the 1.5PN relative order,
of the radiation reaction force \cite{BD88,B93,B97}. Therefore, the
formal post-Newtonian expansion of retarded Green functions is no
longer valid starting at the 4PN order.  We face here a true
difficulty, which is fundamentally linked to the nature of the
post-Newtonian approximation.

The aim of the present article is to resolve the two latter
problems. We shall prove that the post-Newtonian expansion can be {\it
indefinitely} reiterated, while incorporating the correct boundary
conditions satisfied by the wave field at infinity. In particular, we
shall get new insights about the problem of gravitational-radiation
reaction inside an isolated (post-Newtonian) system. To cure the
problem of divergencies we introduce, at any post-Newtonian order, a
generalized solution of the Poisson equation with non-compact support
source, in the form of an appropriate {\it finite part} of the usual
Poisson integral~: namely we regularize the bound at infinity of the
Poisson integral by means of a process of analytic continuation,
analogous to the one already used to regularize the retarded integrals
in Refs. \cite{BD86,B95,B98mult}. Our generalized solution constitutes
a particular (well-defined) solution of the problem; the most general
solution is the sum of that particular solution and the most general
solution of the corresponding homogeneous equation, i.e. the
source-free Laplace equation.  The homogeneous solution should be
regular all over the matter system (we are considering smooth matter
distributions), and we find, after summing up the post-Newtonian
series, that it can be thoroughly written with the help of some
tensorial functions of time $A\mn_L(t)$, where $L=i_1\cdots i_l$
denotes a multi-index with $l$ indices \cite{N4}. At this stage,
considering the post-Newtonian iteration scheme alone, we cannot do
more and therefore we leave the functions $A\mn_L(t)$ unspecified. We
refer to them as some ``radiation-reaction'' functions.

The solution of the problem of the near-zone limitation of the
post-Newtonian expansion resides in the matching of the near-zone
field to the exterior field, a solution of the vacuum equations
outside the source which has been developed in previous works
\cite{BD86,B93} using some post-{\it Minkowskian} and multipolar
expansions. In the case of post-Newtonian sources, the near zone,
i.e. $r\ll\lambda$, covers entirely the source, because the source's
radius itself is such that $a\ll\lambda$. Thus the near zone overlaps
with the exterior zone where the multipole expansion is valid.
Matching together the post-Newtonian and multipolar-post-Minkowskian
solutions in this overlapping region is an application of the method
of matched asymptotic expansions, and has frequently been applied in
the present context, both for radiation-reaction
\cite{BuTh70,Bu71,BD88,B93,B97} and wave-generation
\cite{BD89,DI91a,B95,B98mult} problems.

The exterior multipolar-post-Minkowskian field originally obtained in
Ref. \cite{BD86} depends on some ``multipole-moment'' functions, say
$X\mn_L(t)$ [whose components are associated with some source
multipole moments e.g. $I_L(t)$, $J_L(t)$, $\cdots$], which must be
left unspecified as long as we consider only the external vacuum
solution.  In the work \cite{B98mult}, we have shown that the
multipole moments $X\mn_L(t)$ are entirely determined, up to any
post-Newtonian order, from the requirement of matching to a
post-Newtonian solution.  In the present paper, we shall further show
that the radiation-reaction functions $A\mn_L(t)$, parametrizing the
post-Newtonian solution, are also uniquely fixed, up to any
post-Newtonian order, by the matching. In particular, we shall find
that the latter functions include correctly the contribution of wave
tails, arising at the 4PN order, as determined in
Refs. \cite{BD88,B93,B97}.  We shall also recover by a different
method the result of Ref. \cite{B98mult} concerning the multipole
moments $X\mn_L(t)$.

A comment is in order regarding the possibility of determining the
near-zone field by matched asymptotic expansions up to {\it any}
post-Newtonian order. Indeed the method pre-supposes the existence of
the exterior near-zone for which $a<r\ll\lambda$.  Now if a given
physical system, whose dynamics is described by Newton's theory, emits
gravitational radiation at some Newtonian fundamental wave length
$\lambda_{\mathrm N}$, we expect that when taking into account the
post-Newtonian corrections up to the post-Newtonian order $n$, it will
have a radiation spectrum composed of harmonics between $\sim
2\lambda_{\mathrm N}/n$ and $\sim 2\lambda_{\mathrm N}$.  Indeed this
is the case of the radiation from a binary system moving on a circular
orbit, for which we have $\frac{2}{n+2}\lambda_{\mathrm N}\leq
\lambda_{n\mathrm{PN}}\leq 2\lambda_{\mathrm N}$.  Therefore, if $n$
is large enough, say $n\gtrsim 2\lambda_{\mathrm N}/a$, we expect that
there will be some part of the radiation whose frequency is too high
for the exterior near zone to exist. What we want to say is that the
formulas we shall obtain for the post-Newtonian field of a source ``up
to any order'' are indeed physically valid, strictly speaking, only up
to some finite post-Newtonian order $\sim 2\lambda_{\mathrm N}/a$,
where $a$ is the size of the source; but that, if we consider a source
which is less relativistic, for instance which is obtained by
``slowing down'' our source so that its Newtonian wave length gets
twice its original value (say), the {\it same} post-Newtonian formulas
can then be used for the new source up to approximately twice the
previous post-Newtonian order.

The plan of this paper is as follows. In Section \ref{II} we recall
the construction in Ref. \cite{BD86} of the multipole expansion of the
external field, and we obtain thanks to a result of Ref. \cite{B93}
the near-zone expansion of that external field ready for subsequent
matching. In Section \ref{III} we implement the post-Newtonian
iteration of the inner field inside the matter source, and we find the
far-zone (multipolar) expansion of that post-Newtonian solution, also
ready for matching. In Section \ref{IV} we show that the matching
works up to any post-Newtonian order, and permits the determination of
all the unknowns, in both the external and inner fields.  Finally in
Section \ref{V} we check that our post-Newtonian solution satisfies
the harmonic-coordinate condition as a consequence of the equations of
motion of the source. The technical proofs are relegated to Appendices
\ref{A}, \ref{B} and \ref{C}.

\subsection{Notation for the Einstein field equations}\label{IB}

For the problem at hands let us introduce an asymptotically
Minkowskian coordinate system for which the basic gravitational-wave
amplitude, $h\mn = \sqrt{-g}\, g\mn - \eta\mn$, is divergenceless,
i.e. satisfies the de Donder or harmonic gauge condition $\partial_\mu
h\mn = 0$.  Here, $g\mn$ denotes the contravariant metric (satisfying
$g^{\mu\rho}g_{\rho\nu}=\delta^\mu_\nu$), $g$ is the determinant of
the covariant metric, $g = \mathrm{det}( g_{\mu\nu})$, and $\eta\mn$
represents an auxiliary Minkowskian metric with signature $+2$. With
these definitions the Einstein field equations can be recast into the
d'Alembertian equation

\begin{equation}\label{1}
\Box h\mn = \frac{16\pi G}{c^4} \tau\mn
\;, \end{equation}
where $\Box =
\eta^{\mu\nu}\partial_\mu\partial_\nu=-\frac{1}{c^2}\partial^2/\partial
t^2+\Delta$ is the (flat-spacetime) d'Alembertian operator. The source
term, $\tau\mn$, can rightly be interpreted as the ``effective''
stress-energy pseudo-tensor of the matter and gravitational fields in
harmonic coordinates. It is conserved in the usual sense, and that is
equivalent to the condition of harmonic coordinates~:

\begin{equation}\label{2}
\partial_\mu h\mn =0~~\Longleftrightarrow~~\partial_\mu \tau\mn = 0
\;. \end{equation}
The pseudo-tensor $\tau\mn$ is made of the contribution of the matter
fields, described by a stress-energy tensor $T\mn$, and the
one due to the gravitational field, given by the
gravitational source term $\Lambda\mn$; thus,

\begin{equation}\label{3}
\tau\mn = |g| T\mn+ \frac{c^4}{16\pi
G}\Lambda\mn
\;. \end{equation}
The conservation property (\ref{2}) is equivalent to the
conservation, in the covariant sense, of the matter tensor~:
$\nabla_\mu T\mn =0$. The exact expression of $\Lambda\mn$, taking
into account all the non-linearities of the Einstein field equations,
reads

\begin{eqnarray}\label{4}
\Lambda\mn = &-& h^{\rho\sigma}
\partial^2_{\rho\sigma} h\mn+\partial_\rho h^{\mu\sigma} 
\partial_\sigma h^{\nu\rho} 
+\frac{1}{2}g\mn g_{\rho\sigma}\partial_\lambda h^{\rho\tau} 
\partial_\tau h^{\sigma\lambda} \nonumber\\
&-&g^{\mu\rho}g_{\sigma\tau}\partial_\lambda h^{\nu\tau} 
\partial_\rho h^{\sigma\lambda} 
-g^{\nu\rho}g_{\sigma\tau}\partial_\lambda h^{\mu\tau} 
\partial_\rho h^{\sigma\lambda} 
+g_{\rho\sigma}g^{\lambda\tau}\partial_\lambda h^{\mu\rho} 
\partial_\tau h^{\nu\sigma}\nonumber\\
&+&\frac{1}{8}(2g^{\mu\rho}g^{\nu\sigma}-g^{\mu\nu}g^{\rho\sigma})
(2g_{\lambda\tau}g_{\epsilon\pi}-g_{\tau\epsilon}g_{\lambda\pi})
\partial_\rho h^{\lambda\pi} 
\partial_\sigma h^{\tau\epsilon}
\;. \end{eqnarray}
It is clear from this expression that $\Lambda\mn$ is made of terms
which are at least quadratic in the gravitational-field strength
$h\mn$ and its first and second space-time derivatives.

In this article, we look for the solutions of the field equations
(\ref{1})-(\ref{4}) under the following hypotheses. First, we assume
that the matter tensor $T\mn$ has a spatially compact support,
i.e. can be enclosed into some time-like world tube, say $r\leq a$,
where $r=|{\mathbf x}|$ is the harmonic-coordinate radial
distance. Second, we assume that the matter distribution inside the
source is smooth~: i.e. $T\mn({\mathbf x},t) \in C^\infty
(\Real^4)$. We have in mind a smooth hydrodynamical ``fluid'' system,
without any singularities nor shocks ({\it a priori}), that is
described by some Eulerian-type equations including high relativistic
(post-Newtonian) corrections. In particular, we exclude from the start
any sources containing black holes.  Notice, however, that it makes
sense to apply the formulas derived {\it a priori} only for smooth
matter distributions to systems containing compact objects (including
black holes), described by some sort of point-particle singularities;
see e.g. Refs. \cite{BDI95,WWi96,B96,B98tail,BIJ01}.  Finally, in
order to select the physically sensible solution of the field
equations, we choose some boundary conditions at infinity
corresponding to the famous no-incoming radiation condition. In this
paper, we shall rely on a specific construction of the metric outside
the domain of the source ($r>a$), that was achieved in
Ref. \cite{BD86} under the assumption that the gravitational field has
been independent of time (stationary) in some remote past, in the
sense that $t\leq -{\cal T}~\Longrightarrow~\frac{\partial}{\partial
t}\left[h\mn({\mathbf x},t)\right] = 0$.  This condition is a mean to
impose, by brute force, the no-incoming radiation condition \footnote{However the condition of stationarity in the past, though
much weaker than the actual no-incoming radiation condition, does not
seem to entail any physical restriction on the applicability of the
formalism, even in the case of sources which have always been
radiating.}.

\section{Exterior field}\label{II}

\subsection{Multipolar expansion of the non-linear vacuum field}\label{IIA}

In this section we review some material from Ref. \cite{BD86}
concerning the construction of {\it vacuum} metrics by means of mixed
multipolar and post-Minkowskian (MPM) expansions. The so-called MPM
metrics aim at describing the gravitational field in the region
exterior to a general isolated system. In fact they are mathematically
defined in the open domain $\Real^3_*\times\Real$, i.e. $\Real^4$
deprived from the spatial origin $r\equiv |{\mathbf x}|=0$, but of
course do not agree physically with the real solution when $0<r<a$,
since they are vacuum solutions. For our present purpose the point is
that the most general physically admissible solution of the vacuum
field equations has been obtained in Ref. \cite{BD86} by a specific
construction of the post-Minkowskian solution, say

\begin{equation}\label{5}
h\mn_\mathrm{ext}=\sum_{m=1}^{+\infty}G^mh\mn_{(m)} \;,
\end{equation} whose coefficients are in the form of multipolar
series, or equivalently decompositions in symmetric-trace-free (STF)
products of unit vectors $\hat{n}_L$, that are equivalent to the usual
decomposition in spherical harmonics \cite{N4}~:

\begin{equation}\label{6}
\forall m\geq 1\;,\quad h\mn_{(m)}({\mathbf x},t)=\sum_{l=0}^{+\infty}
\hat{n}_L(\theta,\phi) ~\!h\mn_{(m)L}(r,t)
\;. \end{equation}
The $h\mn_{(m)L}$'s are certain functions of the radial coordinate $r$
and of time $t$.  Inserting the MPM expansion (\ref{5})-(\ref{6}) into
the vacuum field equations (\ref{1})-(\ref{2}) we obtain, at any
post-Minkowskian order $m$,

\begin{eqnarray}\label{7}
\Box h\mn_{(m)} &=& \Lambda\mn_{(m)}\left[h_{(1)}, \cdots, h_{(m-1)}\right]\;,\\
\partial_\mu h\mn_{(m)} &=& 0
\;, \end{eqnarray}
where $\Lambda\mn_{(m)}$ denotes the $m$-th post-Minkowskian piece of
the gravitational source term defined by Eq. (\ref{4}), i.e. in which
we have inserted the previous post-Minkowskian iterations up to the
previous order $m-1$ [with the convention that $\Lambda\mn_{(1)}=0$].
Because Eq. (\ref{4}) is at least quadratic in non-linearities, it is
clear that only the preceding iterations, $\leq m-1$, are necessary at
any post-Minkowskian order $m$.

Now the solution that was obtained in Ref. \cite{BD86} has two main
characteristics. The first one is related to its particular near-zone
structure, which will play a fundamental role in the present
article. Namely, it was proved that each one of the
multipolar-post-Minkowskian coefficients $h\mn_{(m)}$ in Eq. (\ref{5})
-- that we recall are only defined when $r>0$ --, admits a singular
near-zone expansion, i.e. when $r\to 0$, owning the following
structure~:

\begin{equation}\label{8}
\forall N\in\N\;,\quad h\mn_{(m)}({\mathbf x},t) =~ \sum_{l,a,p} {\hat
n}_L r^a (\ln r)^p F\mn_{(m)L,a,p}(t) + R\mn_{(m)N}({\mathbf x},t) \;,
\end{equation} where the multipolar order $l\in\N$, where the powers
of $r$ are such that $a\in\Z$ with $a_\mathrm{min}\leq a\leq N$ (with
$a_{\mathrm{min}}$ a negative integer), and where the powers of $\ln
r$ are $p\in\N$ with $p\leq m-1$.  The maximal divergence when $r\to
0$ occurs for $a_\mathrm{min}$, which depends on the post-Minkowskian
order $m$, and satisfies $a_\mathrm{min}(m)\to -\infty$ when $m\to
+\infty$. Similarly the maximal power of the logarithms,
$p_{\mathrm{max}}(m)=m-1$, tends to infinity with $m$.  The functions
$F\mn_{(m)L,a,p}(t)$ are smooth functions of time, $F\mn_{(m)L,a,p}\in
C^\infty(\Real)$, which are to be computed by means of the algorithm
proposed in Ref. \cite{BD86}, and appear as complicated non-linear
functionals of some more elementary functions parametrizing the
linearized ($m=1$) approximation.  The remainder term in Eq. (\ref{8})
is such that

\begin{equation}\label{9}
R\mn_{(m)N}({\mathbf x},t) ={\cal O}(r^N)\quad\hbox{when $r\to 0$ and $t=$ const}
\;. \end{equation}
The Landau ${\cal O}$-symbol takes its usual meaning. This remainder
admits also some specific differentiability properties (refer to
\cite{BD86} for the details). The gravitational source term
$\Lambda\mn_{(m)}$ admits exactly the same near-zone structure as in
Eq. (\ref{8}) with the exception that $p_\mathrm{max}=m-2$ in this case
(that is, the maximal power of the logarithms increases by one unit
when going from the source to the solution).

The second important characteristic of the MPM solution concerns the
constructive formula which defines it. We find that each one of the
post-Minkowskian coefficients $h\mn_{(m)}$ is explicitly constructed
by means of the following \cite{BD86}~:

\begin{equation}\label{10}
h\mn_{(m)} = \pf\Box^{-1}_\mathrm{Ret}\left[{\widetilde r}^B
\Lambda\mn_{(m)}\right] + \sum^{+\infty}_{l=0} \hat{\partial}_L
\left\{ {1\over r} X^{\mu\nu}_{(m)L} \left(t-\frac{r}{c}\right)
\right\}
\;. \end{equation}
The first term involves a special type of generalized inverse
d'Alembertian operator, built on the standard retarded integral,

\begin{equation}\label{11}
\Box^{-1}_\mathrm{Ret}\left[{\widetilde r}^B \Lambda\mn_{(m)}\right]({\mathbf
x},t) \equiv -\frac{1}{4\pi} \int_{\Real^3} ~\frac{d^3{\mathbf y}}{|{\mathbf
x}-{\mathbf y}|} ~\!|{\widetilde {\mathbf y}}|^B\Lambda_{(m)}\mn({\mathbf y},
t-|{\mathbf x}-{\mathbf y}|/c)
\;, \end{equation}
which extends over the whole three-dimensional space, but inside which a
regularization factor has been ``artificially'' introduced, namely

\begin{equation}\label{12}
{\widetilde r}^B \equiv \left(\frac{r}{r_0}\right)^B
\;, \end{equation}
where $B$ denotes a complex number, $B\in\C$, and $r_0$
represents an arbitrary constant length scale. The indication $\pf$
stands for the {\it finite part} at $B=0$, and means that one should
first compute the Laurent expansion when $B\to 0$ of (the analytic
continuation of) the $B$-dependent integral (\ref{11}), and, second,
pick up the finite part at $B=0$ in that expansion, i.e. the
coefficient of the zero-th power of $B$. The main property of this
generalized retarded operator, that we shall from now on abbreviate as

\begin{equation}\label{13}
\Dlt\big[\Lambda\mn_{(m)}\big] ~\equiv
~\pf\Box^{-1}_{\mathrm{Ret}}~\!\left[{\widetilde r}^B
\Lambda\mn_{(m)}\right] \;, \end{equation} is that, for source terms
$\Lambda\mn_{(m)}$ admitting a near-zone structure of the type
(\ref{8}),

\begin{equation}\label{14}
\Box\left[\Dlt\Lambda\mn_{(m)}\right] = \Lambda\mn_{(m)}
\;. \end{equation}
Because the second term in Eq. (\ref{10}) is a retarded solution of
the {\it source-free} wave equation, we see therefore that
$h\mn_{(m)}$ represents indeed a solution of the wave equation we had
to solve~: $\Box h\mn_{(m)} = \Lambda\mn_{(m)}$. However this is not
sufficient because we have also to solve the harmonic-coordinate
condition (\ref{2}). We shall refer to \cite{BD86} for the definition
of an algorithm which permits to compute, simply from the algebraic
and differential structure of the vacuum field equations, the
necessary form of the second term in Eq. (\ref{10}), in such a way
that the harmonic-coordinate condition be satisfied~: $\partial_\mu
h\mn_{(m)}=0$.  In fact we shall not need, in the following, to be
more precise about the latter term; simply we keep it into the form of
a general retarded solution of the source-free wave equation,
parametrized by some tensorial functions $X\mn_{(m)L}(t)$. We assume
that these functions are STF with respect to the multi-index $L$~:
i.e. $X\mn_{(m)L} \equiv \hat{X}\mn_{(m)L}$, so the multi-derivative
$\hat{\partial}_L$ in Eq. (\ref{10}) is a STF one (see Ref. \cite{N4}
for the notation). The latter construction represents the most general
physical solution of the field equations outside the source
\cite{BD86}.

Let us now proceed with the formal re-summation of the
post-Minkowskian series. That is, once the results
(\ref{8})-(\ref{10}) have been established for any order $m$, we sum
them from $m=1$ up to infinity. In this way we obtain some formulas
which are valid formally for the complete post-Minkowskian series,
and, presumably, could hold true in a more rigorous context of exact
solutions. After summation we shall ``forget'' about the
post-Minkowskian expansion, and consider that the exterior field
$h\mn_\mathrm{ext}$ represents merely the {\it multipole}
decomposition of the actual field $h\mn$ outside the compact support
of the source (nevertheless, it is wise to keep in mind that the
solution came from a formal post-Minkowskian summation). We denote the
multipole decomposition by means of the calligraphic letter ${\cal
M}$. Therefore, our {\it definition} is that the multipole expansion
${\cal M}(h\mn)$ of the field outside the isolated source is nothing
but the external solution constructed previously by means of the MPM
method, and re-summed over the post-Minkowskian index $m$~:

\begin{equation}\label{15}
{\cal M}(h\mn) \equiv h\mn_\mathrm{ext} \;. \end{equation} This
definition is quite legitimate (and rather obvious) because we know
that the MPM metric constitutes the most general solution for the
exterior field. Thus, ${\cal M}(h\mn)$ is a solution of the vacuum
field equations, now considered outside the physical domain of the
source, $r>a$ (while $h\mn_\mathrm{ext}$ had been constructed for any
$r>0$). In that domain we have evidently the numerical equality

\begin{equation}\label{16}
{\cal M}(h\mn) = h\mn\quad\hbox{(when $r>a$)}
\;. \end{equation}
After summation of Eqs. (\ref{8})-(\ref{9}) over $m$, we
get the near-zone structure

\begin{equation}\label{17}
\forall N\in\N\;,\quad {\cal M}(h\mn)  =~ \sum {\hat n}_L r^a (\ln r)^p
F\mn_{L,a,p}(t) + {\cal O}(r^N) 
\;, \end{equation}
in which the functions
$F\mn_{L,a,p}(t)=\sum_{m=1}^{+\infty}G^mF\mn_{(m)L,a,p}(t)$, and where
$a\leq N$ and $p\geq 0$. Notice that there is no lower bound for $a$
because $a_\mathrm{min}(m)\to -\infty$ when $m\to +\infty$; similarly
there is no upper bound for $p$.  Secondly, coming to the constructive
formula (\ref{10}) we obtain

\begin{equation}\label{18}
{\cal M}(h\mn) = \Dlt\left[{\cal M}(\Lambda\mn)\right] +
\sum^{+\infty}_{l=0} \hat{\partial}_L \left\{ {1\over r} X\mn_L
\left(t-\frac{r}{c}\right) \right\} \;, \end{equation} where
$X\mn_L(t)=\sum_{m=1}^{+\infty}G^mX\mn_{(m)L}(t)$. In the following we
shall regard the STF functions $X\mn_L(t)$ as the ``multipole
moments'' of the source, because they describe the physics of the
source as seen from the exterior. We do not need to be more precise at
this point.  Let us simply comment that by imposing the harmonic-gauge
condition (\ref{2}) we find that there are only six components of
these functions which are independent, and this yields the definition
of six independent STF source multipole moments $I_L(t)$, $J_L(t)$,
$\cdots$ (see Ref. \cite{B98mult} for the precise
definition). Furthermore, the multipole-moment functions $X\mn_L(t)$
have already been calculated in terms of the stress-energy tensor of a
post-Newtonian source in Ref. \cite{B98mult}. However we prefer to
leave these functions undetermined because we shall recover their
expressions by means of a somewhat different method, and the agreement
we shall find with the result of Ref. \cite{B98mult} will constitute a
crucial check of our computation.

\subsection{Near-zone expansion of the multipole decomposition}\label{IIB}

In anticipation of the matching we consider next the infinite
near-zone re-expansion, when $r\to 0$, of the multipole expansion
${\cal M}(h\mn)$ determined in Eq. (\ref{18}). We have already
obtained the general structure of that expansion, given by
Eq. (\ref{17}). Let us denote with the help of some overline the {\it
infinite} near-zone expansion (without remainder), whose structure is
therefore given by

\begin{equation}\label{19}
\overline {{\cal M}(h\mn)} =~ \sum {\hat n}_L r^a (\ln r)^p
F\mn_{L,a,p}(t) 
\;, \end{equation} 
where $a\in\Z$ and $p\in\N$ (and, of course, the multipolar
index $l\in\N$). We must be careful at distinguishing the
fully-fledged multipole decomposition ${\cal M}(h\mn)$, which is
defined as soon as $r>0$ and numerically agrees with the exact
solution wherever $r>a$ (in particular when $r\to +\infty$), from its
formal near-zone re-expansion $\overline {{\cal M}(h\mn)}$.  
Later we shall indicate the post-Newtonian expansion by
means of the same overline notation. Indeed the near-zone expansion is really
an expansion when $r/\lambda\to 0$, which is equivalent to an
expansion when $c\to +\infty$, since the wavelength of waves is
$\lambda=cP$ (with $P=$ typical period of the internal motion).  From
the result (\ref{18}) we can write

\begin{equation}\label{20}
\overline{{\cal M}(h\mn)} = \overline{\Dlt\left[{\cal
M}(\Lambda\mn)\right]} + \sum^{+\infty}_{l=0} \hat{\partial}_L \left\{
\frac{\overline{X^{\mu\nu}_L (t-r/c)}}{r} \right\}
\;. \end{equation}
The overline in the second term means that one should expand
the retardations $t-r/c$ when $r/c \to 0$. More
explicitly we have

\begin{equation}\label{21}
\hat{\partial}_L \left\{ \frac{\overline{X^{\mu\nu}_L (t-r/c)}}{r}
\right\} =
\sum_{j=0}^{+\infty}\frac{(-)^j}{c^jj!}\hat{\partial}_L(r^{j-1})
\overset{(j)}{X^{\mu\nu}}_{\!\!\!\!L}(t)
\;, \end{equation}
where the superscript $(j)$ indicates $j$ successive time-derivations.
The main problem is how to treat the first term in
Eq. (\ref{20}). What we essentially want is to know how one can
``commute'' the operations of taking the near-zone expansion and
of applying the retarded integral. In fact, the problem has already been
solved in Ref. \cite{B93}, which succeeded in writing the first term
in Eq. (\ref{20}) as the sum of an ``instantaneous'' operator, 
acting on the near-zone expansion of the source,
and of a particular ``anti-symmetric'' wave (i.e. retarded minus
advanced), solution of the source-free d'Alembertian equation. The
result of Ref. \cite{B93}, Eq. (3.2), reads

\begin{equation}\label{22}
\overline{\Dlt\left[{\cal
 M}(\Lambda\mn)\right]}=\widetilde{\im}\left[\pn{\pmk(\Lambda\mn)}\right]
 - {4G\over c^4} \sum^{+\infty}_{l=0} {(-)^l\over l!} \hat{\partial}_L
 \left\{ \frac{\overline{{\cal R}\mn_L (t-r/c)-{\cal R}\mn_L
 (t+r/c)}}{2r} \right\} \;. \end{equation} For completeness we present
 in Appendix \ref{A} the proof of this result -- a version of it which
 is somewhat improved with respect to that given in
 Ref. \cite{B93}. The first term in Eq. (\ref{22}) involves an
 operator $\widetilde{\im}$, acting on each of the individual terms of
 the formal near-zone expansion whose structure is given by
 Eq. (\ref{19}), and which is essentially defined by the solution of
 the wave equation that is obtained by iterated use of inverse Laplace
 operators, and regularized by means of our $B$-dependent finite part
 procedure. Thus,

\begin{equation}\label{23}
\widetilde{\im}\left[\pn{\pmk(\Lambda\mn)}\right] =
\pf\sum_{k=0}^{+\infty}\left(\frac{\partial}{c\partial t}
\right)^{\!\!2k}\Delta^{-k-1}\left[{\widetilde
r}^B\pn{\pmk(\Lambda\mn)}\right]
\;, \end{equation}
where $\Delta^{-k-1}=(\Delta^{-1})^{k+1}$, and the action of the
inverse Laplacian on the generic term of Eq. (\ref{19}) follows from

\begin{equation}\label{23'}
\Delta^{-1}\left[{\hat n}_L r^{B+a} (\ln r)^p\right] =
\left(\frac{d}{dB}\right)^p\left[\frac{{\hat n}_L
r^{B+a+2}}{(B+a+2-l)(B+a+3+l)}\right] \end{equation} [see also
Eq. (\ref{82}) in Appendix \ref{A}].  The operator $\widetilde{\im}$
plays the central role in the present paper.  It can be regarded as
(the regularization of) the formal post-Newtonian expansion, when
$c\to +\infty$, of the inverse d'Alembert operator, say
$\im=\overline{1/\Box}=\overline{1/(\Delta-\frac{1}{c^2}\partial_t^2)}$. We
can refer to $\im$ as the operator of the instantaneous potentials,
because it acts on the time variable $t$ only through
time-derivations, instead of involving a full integration like for the
operator of the retarded potentials $\Box^{-1}_\mathrm{Ret}$. Notice
that $\im$ is closely related to the operator of the symmetric
potentials,
$\frac{1}{2}\left[\Box^{-1}_{\mathrm{Ret}}+\Box^{-1}_\mathrm{Adv}\right]$;
see Ref. \cite{B93} for discussion and the precise relation between
these operators.  As for the second term in Eq. (\ref{22}), it is made
of an ``anti-symmetric'' wave, which represents in fact a solution of
the d'Alembertian equation that is regular in a neighbourhood of the
origin $r=0$. Its near-zone expansion $r/c\to 0$ is composed only of
terms containing some odd powers of $1/c$~:

\begin{eqnarray}\label{24}
\hat{\partial}_L \left\{ \frac{\overline{{\cal R}\mn_L (t-r/c)-{\cal
 R}\mn_L (t+r/c)}}{2r} \right\} &=&
 -\sum_{i=l}^{+\infty}\frac{\hat{\partial}_L(r^{2i})}{(2i+1)!}
 \frac{\overset{(2i+1)}{{\cal R}\mn}_{\!\!\!\!\!L}
 (t)}{c^{2i+1}}\nonumber\\ &=&
 -\frac{1}{(2l+1)!!}\sum_{k=0}^{+\infty}\Dn{-k}(\hat{x}_L)
 \frac{\overset{(2k+2l+1)}{{\cal R}\mn}_{\!\!\!\!\!\!\!\!L}
 (t)}{c^{2k+2l+1}} \;. \end{eqnarray} See also Eqs. (2.4)-(2.7) in
 Ref. \cite{B93} for alternative forms of the anti-symmetric wave.  In
 the second of Eqs. (\ref{24}) we have introduced the useful object

\begin{equation}\label{24'}
\Dn{-k}({\hat x}_L) = \frac{(2l+1)!!}{(2k)!!(2l+2k+1)!!}r^{2k}{\hat
x}_L \;, \end{equation} which represents the iterated Laplacian
operator $\Dn{-k}$, regularized by means of the $\pf$ process, acting
on ${\hat x}_L$ which denotes the STF projection of the product
$x_L\equiv x_{i_1}\cdots x_{i_l}$ \cite{N4}. [See also Eq. (\ref{127})
for an alternative expression of the same object.]  From
Ref. \cite{B93}, or from Eq. (\ref{77}) in Appendix \ref{A}, we get
the expression of the functions parametrizing the anti-symmetric
waves,

\begin{equation}\label{25}
{\cal R}\mn_L(t) = \pf\int d^3{\mathbf x}~\!|\widetilde{{\mathbf
x}}|^B {\hat x}_L \int_1^{+\infty}dz~\!\gamma_l(z)~\!{\cal
M}(\tau\mn)({\mathbf x},t-z |{\mathbf x}|/c) \;, \end{equation} where
$|\widetilde{{\mathbf x}}|=|{\mathbf x}|/r_0$ [see
Eq. (\ref{12})]. These functions depend on the whole past-history of
the source \cite{NN}. The $z$-integration involves the weighting
function defined by

\begin{equation}\label{26}
\gamma_l(z)=(-)^{l+1}\frac{(2l+1)!!}{2^l l!}(z^2-1)^l
\;. \end{equation}
This function is normalized so that $\int_1^{+\infty} dz~\!\gamma_l(z)
= 1$, where the value of the integral is obtained by analytic
continuation for $l\in\C$ (see Appendix \ref{A}).  As shown in
Ref. \cite{B93} (see notably Section III.D there), the anti-symmetric
waves in Eq. (\ref{22}) are associated with gravitational radiation reaction
effects of non-linear origin. In particular they contain the
contribution of wave tails in the radiation reaction force, which
appears at the 1.5PN order relatively to the lowest-order radiation
damping, i.e. 4PN order in the equations of motion
\cite{BD88}.  To summarize this subsection, we have obtained the
near-zone re-expansion of the multipole expansion ${\cal
M}(h\mn)$ as

\begin{eqnarray}\label{28}
\overline{{\cal M}(h\mn)} &=&
\widetilde{\im}\left[\pn{\pmk(\Lambda\mn)}\right] \nonumber\\ &-&
{4G\over c^4} \sum^{+\infty}_{l=0} {(-)^l\over l!} \hat{\partial}_L \left\{
\frac{\overline{{\cal R}\mn_L (t-r/c)-{\cal R}\mn_L (t+r/c)}}{2r}
\right\}\nonumber\\&+& \sum^{+\infty}_{l=0} \hat{\partial}_L \left\{
\frac{\overline{X^{\mu\nu}_L (t-r/c)}}{r} \right\}
\;. \end{eqnarray}
The functions ${\cal R}\mn_L(t)$ are known from Eq. (\ref{24}), but
the multipole-moments $X\mn_L(t)$ have not yet been specified at this
stage (though they have already been calculated in
Ref. \cite{B98mult}). Therefore, we have succeeded in computing the
near-zone expansion $\overline{{\cal M}(h\mn)}$ as a functional of the
sole unknown constituted by the $X\mn_L$'s; only by matching can these
functions be determined.

\section{Interior field}\label{III}

\subsection{Post-Newtonian expansion of the near-zone field}\label{IIIA}

Up to now, we have solved the Einstein field equations in the vacuum
outside an isolated source ($r>a$), without any reference to the
stress-energy tensor $T\mn$ of the matter source.  Our next task is to
investigate the field equations inside and in the vicinity of the
matter source, and more precisely in the so-called near-zone, or
region for which $r\ll\lambda$, where $\lambda$ is the typical
wavelength of the emitted waves. From now on we restrict attention to
a post-Newtonian source, whose radius is $a\ll\lambda$.  For
post-Newtonian sources the near zone overlaps with the external region
in what we shall refer to as the matching region, for which
$a<r\ll\lambda$. In the matching region both the multipolar expansion
of the exterior field and the post-Newtonian expansion of the inner
field are legitimate.

Let us denote by means of an overline the formal (infinite)
post-Newtonian expansion of the field inside the source's near-zone~:
$\pn{h}\mn$, which is of the form

\begin{equation}\label{29}
 \pn{h}\mn({\mathbf
 x},t,c)=\sum_{n=2}^{+\infty}\frac{1}{c^n}~\!\underset{n}{\pn{h}}\mn({\mathbf
 x},t,\ln c)
\;. \end{equation}
By definition, the $n$-th post-Newtonian coefficient
$\underset{n}{\pn{h}}\mn$ is the factor of the $n$-th power of $1/c$;
however, we know from the structure of the near-zone expansion of the
exterior field [see Eq. (\ref{19})] that the post-Newtonian expansion
will involve also, besides the usual powers of $1/c$, some logarithms
of $c$ (in fact when stating this we are anticipating on the result of
the matching). So the coefficients $\underset{n}{\pn{h}}\mn$ still
depend on $c$ {\it via} the logarithm of $\ln c$, and from
Eq. (\ref{19}) we infer that they are in fact some power series in
$\ln c$. The first appearance of $\ln c$ is at the 4PN order
(i.e. corresponding to a term $\sim\ln c/c^8$ in the equations of
motion) and is associated to the physical effect of wave tails
\cite{BD88}. In Eq. (\ref{29}) we have indicated that the expansion
starts at the level $1/c^2$, but we could be more precise because the
$0i$ component of ${\pn h}\mn$ starts only at the level $1/c^3$, while
the $ij$ component is at least of order $1/c^4$. This does not matter
for our purpose; simply in our iteration we include these
post-Newtonian coefficients as zero~: $\underset{2}{\pn{h}}^{0i}=0$,
and $\underset{2}{\pn{h}}^{ij}=\underset{3}{\pn{h}}^{ij}=0$.  For the
total stress-energy pseudo-tensor (\ref{3}) we have the same type of
expansion,

\begin{equation}\label{30}
 \pn{\tau}\mn({\mathbf
 x},t,c)=\sum_{n=-2}^{+\infty}\frac{1}{c^n}~\!\underset{n}{\pn{\tau}}\mn({\mathbf
 x},t,\ln c)
\;. \end{equation}
The expansion starts with a term of order $c^2$
corresponding to the rest mass-energy of the source (${\pn \tau}\mn$
has the dimension of an energy density). Here we shall always
understand the infinite sums such as (\ref{29})-(\ref{30}) in
the sense of {\it formal} power series, i.e. merely as an ordered
collection of coefficients~:
e.g. $\left(\underset{n}{\pn{h}}\mn\right)_{n\in \N}$.  We do not
attempt to control the mathematical nature of these series.

In this article we make two important assumptions. First, we assume
that the post-Newtonian coefficients $\underset{n}{\pn{h}}\mn$ (and
similarly $\underset{n}{\pn{\tau}}\mn$) are {\it smooth} functions of
space-time,

\begin{equation}\label{31}
\underset{n}{\pn{h}}\mn({\mathbf x},t) ~\in~ C^\infty(\Real^4)
\;. \end{equation} Evidently this comes from our consideration of
regular (smooth) extended matter distributions, described by
$T\mn({\mathbf x},t) \in C^\infty (\Real^4)$, {\it a priori} excluding
black holes or point-particle singularities. Second, we assume that
the structure of the expansion at {\it spatial infinity}, i.e. $r\to
+\infty$ with $t=$ const, is of the type

\begin{equation}\label{32}
\forall N\in\N\;,\quad \underset{n}{\pn{h}}\mn =~ \sum {\hat n}_L r^a (\ln r)^p
F\mn_{L,n,a,p}(t) + {\cal O}\left(\frac{1}{r^N}\right) 
 \end{equation} 
(and similarly for each $\underset{n}{\pn{\tau}}\mn$). On purpose we
have written an expansion which is very similar to the one in
Eq. (\ref{17}), because as we shall see the functions $F_{L,n,a,p}(t)$
will be equal to the post-Newtonian coefficients of the functions
$F_{L,a,p}(t)$ appearing in Eq. (\ref{17}). However it is important to
realize that in contrast to Eq. (\ref{17}) which is a near-zone
expansion [{\it cf}. the remainder ${\cal O}(r^N)$], the expansion
written in Eq. (\ref{32}) is a {\it far-zone} one, with remainder
${\cal O}(1/r^N)$. It would have been clearer to write the latter
expansion with some $(\ln r)^p/r^b$ with $b=-a$, but since we are
going to show, from the method of matched asymptotic expansions, that
the infinite far-zone expansion (ignoring the remainder) is actually
the {\it same} as the infinite near-zone expansion, it is better to
write it in this form, with the range of the powers of $r$ in
Eq. (\ref{32}) being $-a\leq N$ instead of $a\leq N$ in
Eq. (\ref{17}). In doing so we are again anticipating on the result of
the matching.  Finally, we assume that, at any given post-Newtonian
order $n$, the maximal divergency of the far-zone expansion (\ref{32})
is finite, i.e. there exists some $a_\mathrm{max}(n)\in\N$ such that
$a\leq a_\mathrm{max}(n)$.

Next we perform the iteration of the post-Newtonian field (\ref{29})
up to any order.  Our strategy consists of finding the general
post-Newtonian solution of the relaxed Einstein field equation
(\ref{1}).  This solution will depend on some arbitrary
``homogeneous'' solutions, in the form of harmonic solutions solving
the source-free d'Alembertian equation (in a perturbative
post-Newtonian sense).  In a second stage we shall obtain these
harmonic solutions by imposing the matching to the external multipolar
field obtained in Section \ref{II}. Finally we shall check that our
post-Newtonian solution is divergenceless, i.e. it satisfies the
harmonic-coordinate condition (\ref{2}), in consequence of the
conservation of the stress-energy pseudo tensor $\tau\mn$. Notice that
we do not try to incorporate into the post-Newtonian series the
boundary conditions at infinity ({\it viz} the no-incoming radiation
condition). Indeed this is impossible at the level of the
post-Newtonian expansion considered alone, because its validity is
limited to the near-zone. Even if we define an ``improved''
post-Newtonian series by considering some {\it retarded} integrals
that are formally expanded when $c\to +\infty$ as series of
Poisson-like integrals \cite{AD75}, we ultimately end up with an
inconsistency, because the Poisson-like integrals are some
local-in-time functionals, depending on the source only at the current
time $t$, and we know that the post-Newtonian field starts to depend
on the whole past history of the source from the 4PN order
\cite{BD88,B93,B97}. Therefore, we do not follow this route in the
present paper, and, instead, we incorporate into the post-Newtonian
series the boundary conditions concerning the wave field at infinity
by means of the matching equation.

We insert the post-Newtonian ansatz (\ref{29})-(\ref{30}) into the
``relaxed'' Einstein field equation (\ref{1}), and equate together
the powers of $1/c$. The result is an infinite set of Poisson-type
equations~:

\begin{equation}\label{33}
\forall n\geq 2\;,\qquad\Delta\underset{n}{\pn{h}}\mn =16\pi
G~\!\underset{n-4}{\pn{\tau}\mn}+\partial_{t}^{2}\underset{n-2}{\pn{h}\mn}
\;. \end{equation}
Evidently, the second term comes from the split of the d'Alembertian
operator into a Laplacian and a second time derivative~:
$\Box=\Delta-\frac{1}{c^2}\partial_t^2$; the time derivative
$\partial_0=\frac{1}{c}\partial_t$ is smaller than the spatial
gradient $\partial_i$ by a factor $1/c$ -- this is the basic tenet of
the approximation. When $n=2$ and $n=3$ the second term in
Eq. (\ref{33}) is zero, which we take into account by assuming that
$\underset{0}{\pn{h}}\mn=\underset{1}{\pn{h}}\mn=0$.  We proceed by
induction, i.e. we fix some post-Newtonian order $n$, assume that we
succeeded in constructing the sequence of previous coefficients
$\underset{2}{\pn{h}}\mn, \cdots, \underset{n-1}{\pn{h}\mn}$, and from
this we infer the next-order coefficient $\underset{n}{\pn{h}}\mn$.

The most general solution consists of the sum of a particular solution
and of the most general admissible solution of the homogeneous
equation, which is simply the source-free Laplace equation. Let us
first find a particular solution. We recalled in the introduction that
the usual Poisson integral cannot be used to define a
solution, because the bound at infinity becomes rapidly divergent when
going to higher and higher post-Newtonian orders.  Fortunately, thanks
to our two assumptions (\ref{31}) and (\ref{32}), we shall be able to
define a {\it generalized} notion of Poisson integral, in a way
similar to our previous definition of a retarded integral operator in
Eq. (\ref{13}).  That generalized Poisson integral will constitute an
appropriate solution of the post-Newtonian equation. For any source
term like $\underset{n}{\pn{\tau}}\mn$ which is at once smooth,
Eq. (\ref{31}), and admits a far-zone expansion of the type (\ref{32})
[note that Eqs. (\ref{31})-(\ref{32}) hold for
$\underset{n}{\pn{h}}\mn$ as well as for $\underset{n}{\pn{\tau}}\mn$],
we multiply it by the same regularization factor as in Eq. (\ref{13}),
and then apply the standard Poisson integral. The result,

\begin{equation}\label{34}
\Delta^{-1}\left[{\widetilde
r}^B\underset{n}{\pn{\tau}}\mn\right]({\mathbf x},t) = -\frac{1}{4\pi}
\int_{\Real^3} ~\frac{d^3{\mathbf y}}{|{\mathbf x}-{\mathbf y}|}
~\!|{\widetilde {\mathbf y}}|^B~\!\underset{n}{\pn{\tau}}\mn({\mathbf
y},t) \;, \end{equation} where $|{\widetilde {\mathbf y}}|^B\equiv
|{\mathbf y}/r_0|^B$, defines a certain function of $B\in\C$. The
well-definiteness of that integral heavily relies on the behaviour at
the bound at infinity.  There is no problem with the vicinity of the
origin because of the smoothness of the integrand.  From the
asymptotic expansion (\ref{32}), with $a\leq a_\mathrm{max}$ [recall
that $a_\mathrm{max}=a_\mathrm{max}(n)$], we find that the integral
converges at infinity when $\Re (B)<-a_\mathrm{max}-2$.  Next we can
prove that the latter function of $B$ generates a (unique) analytic
continuation down to a neighbourhood of the origin $B=0$, except at
$B=0$ itself, at which value it admits a Laurent expansion with
multiple poles up to some finite order. More details are given in
Appendix \ref{B}.  Then, we consider the Laurent expansion of that
function when $B\to 0$ and pick up the finite part, or coefficient of
the zero-th power of $B$, of that expansion. This defines our
generalized Poisson integral~:

\begin{equation}\label{35}
\Dt\big[~\!\underset{n}{\pn{\tau}}\mn\big] \equiv
\pf\Delta^{-1}\left[{\widetilde r}^B\underset{n}{\pn{\tau}}\mn\right]
\;. \end{equation}
The finite-part symbol $\pf$ has exactly the same meaning as in
Eq. (\ref{13}).  However, notice that in contrast to Eq. (\ref{13})
where the regularization factor ${\widetilde r}^B$ dealt with the
singularity when $r\to 0$, and hence supposes initially that $\Re (B)$
is a large positive number, in Eq. (\ref{35}) the regularization
concerns the behaviour of the integral when $r\to +\infty$, and so one
must start with the situation where $\Re (B)$ is a large {\it
negative} number.  The main properties of our generalized Poisson
operator is that it solves the Poisson equation,

\begin{equation}\label{36}
\Delta\left[\Dt~\!\underset{n}{\pn{\tau}}\mn\right] =
\underset{n}{\pn{\tau}}\mn \;, \end{equation} and that the solution
$\Dt~\!\underset{n}{\pn{\tau}}\mn$ owns the same properties as the
source $\underset{n}{\pn{\tau}}\mn$, i.e. the smoothness,
Eq. (\ref{31}), and the particular far-zone expansion given by
Eq. (\ref{32}).  These facts are proved in Appendix
\ref{B}. Therefore, we have found a {\it particular} solution of the
Poisson equation, and, furthermore, this solution can be iterated at
will, because the operator $\Dt$ keeps the same properties from the
source to the corresponding solution. Quite naturally we denote the
iterated Poisson operator $\Dn{-k-1}\equiv
\left(\Dn{-1}\right)^{k+1}$; it is not difficult to show that

\begin{equation}\label{37}
\Dn{-k-1} [~\!\underset{n}{\pn{\tau}}\mn]({\mathbf x},t) = -\frac{1}{4\pi}
~\!\pf\int_{\Real^3} ~d^3{\mathbf y}~\!\frac{|{\mathbf x}-{\mathbf y} |^{2k-1}
}{(2k)!}~\!|\widetilde{{\mathbf y}}|^B\underset{n}{\pn{\tau}}\mn({\mathbf
y},t)
\;. \end{equation}
From that integral we obtain the operator of 
the ``instantaneous'' potentials exactly in the
same way as in Eq. (\ref{23}), but now acting on post-Newtonian
coefficients such as $\underset{n}{\pn{\tau}}\mn$, i.e. satisfying
both Eqs. (\ref{31}) and (\ref{32})~:

\begin{equation}\label{38}
\widetilde{\im}[~\!\underset{n}{\pn{\tau}}\mn] =
\sum_{k=0}^{+\infty}\left(\frac{\partial}{c\partial
t}\right)^{\!\!2k}\Dn{-k-1}[~\!\underset{n}{\pn{\tau}}\mn]
\;. \end{equation} It is clear that we have a particular solution of
d'Alembert's equation~:

\begin{equation}\label{39}
\Box\left[\widetilde{\im}~\!\underset{n}{\pn{\tau}}\mn\right] =
\underset{n}{\pn{\tau}}\mn
\;. \end{equation}
We can check that the definition we have proposed in
Eqs. (\ref{23})-(\ref{23'}) is a particular case of the more general
definition (\ref{37})-(\ref{38}). Indeed, if we apply the formulas
(\ref{37})-(\ref{38}) to one of the terms composing the ``far-zone''
expansion of the post-Newtonian coefficient, i.e. ${\hat n}_L r^a (\ln
r)^p F(t)$, we get the same result as the one resulting from
Eqs. (\ref{23})-(\ref{23'}).

By means of the Poisson operator $\Dt$ so constructed we first
find a {\it particular} solution of Eq. (\ref{33})~:

\begin{equation}\label{40}
\left(\underset{n}{\pn{h}}\mn\right)_\mathrm{part}=16\pi
G~\!\Dt\underset{n-4}{\pn{\tau}\mn}
+\partial_{t}^{2}\Dt\underset{n-2}{\pn{h}\mn}
\;. \end{equation}
To this solution we add the most general solution of the homogeneous
Laplace equation. It can be written, using the STF language, as the
sum of two multipolar series, one of them being of the type ${\hat
x}_L$, that is regular at the origin $r=0$, the other one being like
$\hat{\partial}_L(1/r)$, i.e. regular ``at infinity'' $r\to +\infty$
(see Ref. \cite{N4} for the notation). Imposing the smoothness
condition (\ref{31}) for the post-Newtonian field we discard the
second type $\sim\hat{\partial}_L(1/r)$, and retain as the only
admissible homogeneous solution the first type $\sim{\hat
x}_L$. Therefore, we find that there must exist some STF-tensorial
functions of time, say $\underset{\!\!\!n}{B_L\mn}\!(t)$, such that

\begin{equation}\label{41}
\left(\underset{n}{\pn{h}}\mn\right)_{\mathrm{hom}}
=\sum_{l=0}^{+\infty}\underset{\!\!\!n}{B_L\mn}\!(t)~\!{\hat
x}_L \;. \end{equation} The functions
$\underset{\!\!\!n}{B_L\mn}\!(t)$ will be associated with the reaction
of the field onto the source, and depend on which boundary conditions
are to be imposed on the gravitational field at infinity.  The most
general solution for the $n$-th post-Newtonian coefficient thus reads

\begin{equation}\label{42}
\underset{n}{\pn{h}}\mn=\left(\underset{n}{\pn{h}}\mn\right)_{\mathrm{part}}
+\left(\underset{n}{\pn{h}}\mn\right)_\mathrm{hom}
\;. \end{equation} It is now trivial to iterate the process. We
substitute for $\underset{n-2}{\pn{h}\mn}$ in the right-hand-side of
Eq. (\ref{40}) the same expression but with $n$ replaced by $n-2$, and
similarly descend untill we stop at either one of the coefficients
$\underset{0}{\pn{h}}\mn=0$ or $\underset{1}{\pn{h}}\mn=0$. At this
point $\underset{n}{\pn{h}}\mn$ is expressed in terms of the
``previous'' $\underset{m}{\pn{\tau}}\mn$'s and
$\underset{\!\!\!m}{B_L\mn}$'s, with $m\leq n-2$, i.e.

\begin{equation}\label{43}
\underset{n}{\pn{h}}\mn=16\pi G
\sum_{k=0}^{[\frac{n}{2}]-1}\partial_{t}^{2k}\Dn{-k-1}
[\underset{n-4-2k}{\pn{\tau}\mn}]
+\sum_{l=0}^{+\infty}~\!\sum_{k=0}^{[\frac{n}{2}]-1}
\overset{(2k)}{\underset{n-2k}{B_L\mn}}\!(t)~\!\Dn{-k}({\hat
x}_L)
\;. \end{equation} 
Here $[n/2]$ denotes the integer part of $n/2$; $\partial_{t}^{2k}$
means the $2k$-th partial time derivative $(\partial/\partial t)^{2k}$
and the superscript $(2k)$ the $2k$-th total time derivative; the
operator $\Dn{-k-1}$ is the one defined by Eq. (\ref{37}); the object
$\Dn{-k}({\hat x}_L)$ has already been introduced in Eq. (\ref{24'}).
Once we have the result (\ref{43}), we ``re-sum'' it from $n=2$ up to
infinity. After commuting the summations over $n$ and $k$ we arrive at

\begin{equation}\label{44}
\pn{h}\mn=\frac{16\pi G}{c^4} \widetilde{\im}\left[\pn{\tau}\mn\right]
+\sum_{l=0}^{+\infty}~\!\sum_{k=0}^{+\infty}\frac{1}{c^{2k}}
\overset{(2k)}{B_L\mn}(t)~\!\Dn{-k}({\hat
x}_L)
\;, \end{equation} 
where we have recognized the operator of the ``instantaneous''
potentials as defined by Eq. (\ref{38}), and where the functions
$B_L\mn$ read

\begin{equation}\label{45}
B_L\mn(t) =
\sum_{n=2}^{+\infty}\frac{1}{c^n}\underset{\!\!\!n}{B_L\mn}(t)
\;. \end{equation}
A more compact alternative form is

\begin{equation}\label{46}
\pn{h}\mn=\frac{16\pi G}{c^4} \widetilde{\im}\left[\pn{\tau}\mn\right]
+\sum_{l=0}^{+\infty}~\!\Delta~\widetilde{\im}\left[B_L\mn(t)~\!{\hat
x}_L\right]
\;. \end{equation} 
Actually the latter forms are not the best for our purpose. Since the
first term in Eqs. (\ref{44}) or (\ref{46}) is a particular solution
of the d'Alembert equation [see Eq. (\ref{39})], the second term is
necessarily equal to (the near-zone re-expansion of) a homogeneous
solution of the source-free wave equation, and most importantly a
regular solution at it. So it should be in the form of some
anti-symmetric multipolar waves~: retarded minus advanced. Indeed,
this readily follows from the second equality in Eq. (\ref{24}). We
introduce a new definition $A_L\mn(t)$ by posing

\begin{equation}\label{48}
B_L\mn(t) = -\frac{\overset{(2l+1)}{A_L\mn}(t)}{c^{2l+1}(2l+1)!!}  \;,
\end{equation} where the $l$-dependent factor is chosen to match with
Eq. (\ref{24}). [Because of our assumption of stationarity in the
past, $t\leq -{\cal T}$, the relation (\ref{48}) determines
$A_L\mn(t)$ up to a constant. However it is clear that this constant
will cancel out in the anti-symmetric wave in Eq. (\ref{49}).] In
terms of this definition, we find the final result of this section,

\begin{equation}\label{49}
\pn{h}\mn=\frac{16\pi G}{c^4} \widetilde{\im}\left[\pn{\tau}\mn\right]
+ \sum^{+\infty}_{l=0} \hat{\partial}_L \left\{ \frac{\overline{A\mn_L
(t-r/c)-A\mn_L (t+r/c)}}{2r} \right\}
\;, \end{equation} 
where we recall that the overline means the post-Newtonian or
equivalently near-zone expansion [see Eq. (\ref{24})].  For the time
being we shall refer to the $A_L\mn(t)$'s as the {\it
radiation-reaction} functions.

\subsection{Multipole expansion of the post-Newtonian solution}\label{IIIB}

In the previous section we obtained the general solution for the
post-Newtonian expansion in the form (\ref{49}), and parametrized by
some (for the moment) unknown radiation-reaction functions
$A_L\mn(t)$.  To arrive at this we made an assumption concerning the
particular structure for the {\it far-zone} expansion, at spatial
infinity, of the post-Newtonian coefficients~: Eq. (\ref{32}). Here we
shall denote the corresponding infinite expansion (without remainder
term) by means of the same calligraphic letter ${\cal M}$ as used to
denote the multipole expansion, because the far-zone expansion of the
post-Newtonian coefficients is equivalent to a multipolar
decomposition. From Eq. (\ref{32}) we have

\begin{equation}\label{50}
{\cal M}(\underset{n}{\pn{h}}\mn) =~ \sum {\hat n}_L r^a (\ln r)^p
F\mn_{L,n,a,p}(t)
\;. \end{equation} 
So, summing up the post-Newtonian series,

\begin{equation}\label{51}
{\cal M}(\pn{h}\mn) =~ \sum {\hat n}_L r^a (\ln r)^p
F\mn_{L,a,p}(t)
\;, \end{equation} 
where the functions involved are
$F_{L,a,p}(t)=\sum_{n=2}^{+\infty}\frac{1}{c^n}F_{L,n,a,p}(t)$.  As we
can see, the far-zone expansion that we have just postulated is
exactly the same, with the same functions $F_{L,a,p}(t)$, as the
near-zone expansion we had previously written in Eq. (\ref{19}).  This
equality is already the matching equation between the near-zone
expansion of the multipolar field, $\overline {{\cal M}(h\mn)}$, and
the multipolar-far-zone expansion of the post-Newtonian field, ${\cal
M}(\pn{h}\mn)$, whose consequences will be investigated in Section
\ref{IV}.

The fundamental result which is needed for computing the far-zone
expansion of the post-Newtonian series concerns the expansion of the
generalized integral operator $\widetilde{\im}$ acting on the
post-Newtonian source $\pn{\tau}\mn$. More precisely, we are
interested in knowing under which conditions one can commute
$\widetilde{\im}$ with the operation ${\cal M}$ of taking the far-zone
expansion.  Clearly, the two operations can be commuted at the price
of adding some homogeneous solution of the d'Alembert equation. We
prove in Appendix \ref{C} that the latter homogeneous solution is made
of multipolar waves of the {\it symmetric} type, i.e. retarded {\it
plus} advanced. We obtain

\begin{equation}\label{52}
{\cal M}\left(\widetilde{\im}\left[\pn{\tau}\mn\right]\right) =
 \widetilde{\im}[{\cal M}(\pn{\tau}\mn)] - \frac{1}{4\pi}
 \sum^{+\infty}_{l=0} {(-)^l\over l!} \hat{\partial}_L \left\{
 \frac{\overline{{\cal F}\mn_L (t-r/c)+{\cal F}\mn_L (t+r/c)}}{2r}
 \right\}
\;. \end{equation} 
Here the overline notation has the same meaning as in Section
\ref{II}~: this is the Taylor expansion when $r\to 0$, but that
expansion should be considered, {\it a posteriori}, as an expansion
when $r\to +\infty$. That is, our notation means

\begin{eqnarray}\label{53}
\hat{\partial}_L \left\{ \frac{\overline{{\cal F}\mn_L (t-r/c)+{\cal
 F}\mn_L (t+r/c)}}{2r} \right\} &=&
 \sum_{i=0}^{+\infty}\frac{\hat{\partial}_L(r^{2i-1})}{(2i)!}\frac{\overset{(2i)}{{\cal
 F}\mn}_{\!\!\!L} (t)}{c^{2i}}\nonumber\\ &=&
 \sum_{i=0}^{+\infty}\Dn{-i}\left[\hat{\partial}_L\!\left(r^{-1}\right)\right]\frac{\overset{(2i)}{{\cal
 F}\mn}_{\!\!\!L} (t)}{c^{2i}} \;, \end{eqnarray} [see also
 Eq. (\ref{123})] where the right-hand-sides are to be considered as
 some expansions at spatial infinity, of the general type given by
 Eq. (\ref{51}). The functions ${\cal F}^{\mu\nu}_L(t)$ parametrizing
 these symmetric waves are STF and explicitly given by

\begin{equation}\label{53'}
{\cal F}\mn_L (t) = \sum_{j=0}^{+\infty}\frac{1}{c^{2j}} \pf\int
d^3{\mathbf x}~\!|\widetilde{{\mathbf
x}}|^B\Dn{-j}\left[\hat{x}_L\right]\partial_t^{2j}\pn{\tau}\mn({\mathbf
x},t)
\;, \end{equation}
where $\Dn{-j}\left[\hat{x}_L\right]$ is given by Eq. (\ref{24'}). See
the proof in Appendix \ref{C}. An alternative form reads as

\begin{equation}\label{54}
{\cal F}^{\mu\nu}_L (t) = \pf\int d^3 {\mathbf x}~\! |\widetilde{\mathbf x}|^B
 {\hat x}_L \, \overline{\int^1_{-1} dz ~\!\delta_l(z)~\!{\overline \tau^{\mu\nu}} 
 ({\mathbf x}, t-z|{\mathbf x}|/c)} 
\;, \end{equation} 
where the integration over the $z$-dependent cone $t-z|{\mathbf
y}|/c$ involves a weighting function $\delta_l(z)$ that is closely
related to the function $\gamma_l(z)$ introduced in Eq. (\ref{26})~:

\begin{equation}\label{55}
\delta_l (z) = {(2l+1)!!\over 2^{l+1} l!} (1-z^2)^l = -\frac{1}{2}\gamma_l (z) 
\;, \end{equation} 
and whose integral is normalized to one~: $\int^1_{-1} dz~\!\delta_l
(z) = 1$ \footnote{The normalization for the function $\delta_l(z)$ is
consistent with that of the function $\gamma_l(z)$~: $\int_1^{+\infty}
dz~\!\gamma_l (z) = 1$, owing to the fact that the integral
$\int_{-\infty}^{+\infty}dz~\!(1-z^2)^l$ is zero by complex analytic
continuation in $l\in\C$.}. The function $\delta_l(z)$ approaches the Dirac
delta-function in the limit of large $l$~: $\lim_{~l \to
+\infty}\delta_l (z)=\delta (z)$. In Eq. (\ref{54}) we have indicated
by means of an overline the fact that this expression is valid only in
a sense of post-Newtonian expansion. Note that because the latter
post-Newtonian expansion is ``even'', containing only even powers of
$1/c$, one can replace the argument $t-z|{\mathbf y}|/c$ inside ${\cal
F}^{\mu\nu}_L(t)$ equivalently by $t+z|{\mathbf y}|/c$.

Finally, thanks to Eqs. (\ref{52})-(\ref{55}), we are in a
position to write the infinite multipolar-far-zone expansion of the
post-Newtonian solution as

\begin{eqnarray}\label{57}
{\cal M}(\pn{h}\mn) &=& \frac{16\pi G}{c^4}\widetilde{\im}[{\cal
 M}(\pn{\tau}\mn)]\nonumber\\ &-& \frac{4G}{c^4} \sum^{+\infty}_{l=0}
 {(-)^l\over l!} \hat{\partial}_L \left\{ \frac{\overline{{\cal
 F}\mn_L (t-r/c)+{\cal F}\mn_L (t+r/c)}}{2r} \right\}\nonumber\\ &+&
 \sum^{+\infty}_{l=0} \hat{\partial}_L \left\{ \frac{\overline{A\mn_L
 (t-r/c)-A\mn_L (t+r/c)}}{2r} \right\} \;. \end{eqnarray} We recall
 that the radiation-reaction functions $A_L\mn(t)$ are still
 undetermined at this stage. The symmetric and anti-symmetric waves
 are given by Eqs. (\ref{53}) and (\ref{24}) respectively, considered
 here as infinite far-zone expansions.

\section{Matching}\label{IV}

In Section \ref{IIA}, we found the most general expression for the
multipolar expansion ${\cal M}(h\mn)$, satisfying the no-incoming
radiation condition, in terms of some unknown ``multipole-moment''
STF-functions $X\mn_L(t)$ [see Eq. (\ref{18})]. On the other hand, in
Section \ref{IIIA}, we obtained the most general solution for the
post-Newtonian expansion $\pn{h}\mn$, as parametrized by a set of
unknown ``radiation-reaction'' STF-functions $A\mn_L(t)$ [see
Eq. (\ref{49})]. We are now imposing the matching condition

\begin{equation}\label{58}
\overline{{\cal M}(h\mn)} \equiv {\cal M}(\pn{h}\mn)
\;. \end{equation} 
In fact we have already postulated this equation when writing that the
two formal expansions (\ref{19}) and (\ref{51}) are the same.  Recall
that the matching equation (\ref{58}) results from the numerical
equality ${\cal M}(h\mn) = \pn{h}\mn$, verified in the exterior
near-zone~: $a<r\ll\lambda$. It is physically justified only for
post-Newtonian sources, for which the exterior near-zone exists.  The
matching equation is actually a {\it functional} identity, i.e. true
$\forall ({\mathbf x},t) \in \Real^3_* \times \Real$; it identifies,
{\it term-by-term}, two asymptotic singular expansions, each of them
being formally taken outside its own domain of validity. In the
present context, the matching equation insists that the infinite {\it
near-zone} expansion, $r\to 0$, of the exterior multipolar field is
identical to the infinite {\it far-zone} expansion, $r\to +\infty$, of
the inner post-Newtonian field. Let us show now that Eq. (\ref{58})
permits determining all the unknowns of the problem~: i.e., at once,
the multipole moments $X\mn_L$ {\it and} the radiation-reaction
functions $A\mn_L$. In particular we find that the multipole moments
$X\mn_L$ are in agreement with the earlier result derived in
Ref. \cite{B98mult}.

For the sake of clarity we re-state here the two results we
reached for the two sides of Eq. (\ref{58}). The left-hand-side was
obtained in Eq. (\ref{28})~:

\begin{eqnarray}\label{59}
\overline{{\cal M}(h\mn)} &=&
\widetilde{\im}\left[\pn{\pmk(\Lambda\mn)}\right]\nonumber\\ &-&
{4G\over c^4} \sum^{+\infty}_{l=0} {(-)^l\over l!} \hat{\partial}_L
\left\{ \frac{\overline{{\cal R}\mn_L (t-r/c)-{\cal R}\mn_L
(t+r/c)}}{2r} \right\}\nonumber\\ &+& \sum^{+\infty}_{l=0}
\hat{\partial}_L \left\{ \frac{\overline{X^{\mu\nu}_L (t-r/c)}}{r}
\right\}
\;, \end{eqnarray}
in which the functions ${\cal R}^{\mu\nu}_L$, which come from the
non-linearities of the field equations in vacuum, are known from
Eq. (\ref{25}). The right-hand-side of the matching equation was found
in Eq. (\ref{57})~:

\begin{eqnarray}\label{60}
{\cal M}(\pn{h}\mn) &=& \frac{16\pi G}{c^4}\widetilde{\im}[{\cal
 M}(\pn{\tau}\mn)]\nonumber\\ &-& \frac{4G}{c^4} \sum^{+\infty}_{l=0}
 {(-)^l\over l!} \hat{\partial}_L \left\{ \frac{\overline{{\cal
 F}\mn_L (t-r/c)+{\cal F}\mn_L (t+r/c)}}{2r} \right\}\nonumber\\ &+&
 \sum^{+\infty}_{l=0} \hat{\partial}_L \left\{ \frac{\overline{A\mn_L
 (t-r/c)-A\mn_L (t+r/c)}}{2r} \right\}
\;. \end{eqnarray} 
Here, the functions ${\cal F}\mn_L$, which depend on the matter and
gravitational content of the post-Newtonian source, take the definite
expression given by Eqs. (\ref{53'})-(\ref{54}).

Comparing the equations (\ref{59}) and (\ref{60}), we readily discover
that they share an obvious common term, that is the first one. Indeed,
we manifestly have

\begin{equation}\label{61}
\widetilde{\im}\left[\pn{\pmk(\Lambda\mn)}\right] =
\widetilde{\im}[{\cal M}(\pn{\Lambda}\mn)] = \frac{16\pi
G}{c^4}\widetilde{\im}[{\cal M}(\pn{\tau}\mn)] \;. \end{equation} The
first equality comes from the matching equation, as applied to the
gravitational source term $\Lambda\mn$, and the second equality comes
from the fact that the matter tensor $T\mn$ has a compact support, so
that ${\cal M}(T\mn)=0$. Hence the two first terms in Eqs. (\ref{59})
and (\ref{60}) match together.  This is a somewhat remarkable fact,
because most of the complexity of the Einstein field equations is
actually contained into these terms, either $\im\big[~\!\overline
{{\cal M}(\Lambda\mn)}~\!\big]$ for the external field or $\frac{16\pi
G}{c^4}\im\left[{\cal M}(\overline \tau\mn)\right]$ for the inner
one. But for doing the matching we don't need all this complexity;
these two terms match and therefore are to be identified. Notice also
that this is a non-trivial result, since the two sides of
Eq. (\ref{58}) strongly depend on the yet unknown functions $A\mn_L$
and $X\mn_L$, which enter the latter two terms in a very intricate
way, coupled together as they are via the non-linearities of the field
equations.  Nevertheless, the matching equation tells us that these
terms must be rigorously identical.

As soon as we have noticed that the first terms in Eqs. (\ref{59}) and
(\ref{60}) are equal, we can compare the other ones, and because the
retarded and advanced waves have some different structures, they must
be matched independently, so we get {\it two} relations to be
satisfied. We find that these are solved if and only if the
multipole-moments in the exterior field {\it and} the
radiation-reaction functions in the inner field are given by

\begin{eqnarray}\label{62}
X\mn_L(t) &=& -\frac{4G}{c^4}\frac{(-)^l}{l!}{\cal
F}\mn_L(t)\label{62a}\;,\\ A\mn_L(t) &=&
-\frac{4G}{c^4}\frac{(-)^l}{l!}\Big[{\cal F}\mn_L(t) + {\cal
R}\mn_L(t)\Big]\label{62b} \;. \end{eqnarray} Therefore, both the
multipole-moments and the radiation-reaction terms are determined as
some explicit functionals of the pseudo-tensor $\tau\mn$ and nothing
else. (Actually we could add any constant to the definition of
$A\mn_L(t)$, but this is physically irrelevant because the constant
disappears from the anti-symmetric waves; see also Ref. \cite{NN}.)

Finally, by way of summary of the results, we take back the latter
expressions and fill in the external and inner fields, which are then
entirely determined as coming from a unique solution of the Einstein
field equations in harmonic coordinates, valid everywhere inside and
outside the source.  The exterior field is

\begin{equation}\label{63}
{\cal M}(h\mn) = \Dlt\left[{\cal M}(\Lambda\mn)\right] - {4G\over c^4}
 \sum^{+\infty}_{l=0} {(-)^l\over l!} \hat{\partial}_L \left\{
 \frac{{\cal F}\mn_L (t-r/c)}{r}\right\}
\;, \end{equation}
where the multipole moments are given in terms of the post-Newtonian
expansion of the stress-energy pseudo-tensor by

\begin{eqnarray}\label{64}
{\cal F}^{\mu\nu}_L (t) &=& \pf\int d^3 {\mathbf y}~\! |\widetilde{\mathbf y}|^B
 {\hat y}_L \, \overline{\int^1_{-1} dz ~\!\delta_l(z)~\!{\overline
 \tau^{\mu\nu}} ({\mathbf y}, t-z|{\mathbf y}|/c)}\nonumber\\
&=& \sum_{j=0}^{+\infty}\frac{1}{c^{2j}} \pf\int
d^3{\mathbf y}~\!|\widetilde{{\mathbf
y}}|^B\Dn{-j}\left[\hat{y}_L\right]\partial_t^{2j}\pn{\tau}\mn({\mathbf
y},t)
\;. \end{eqnarray}
This result is in perfect agreement with the multipole decomposition
of the exterior field obtained in Ref. \cite{B98mult} [see
Eqs. (3.13)-(3.14) there]. On the other hand, the inner post-Newtonian
field is given by

\begin{equation}\label{65}
\pn{h}\mn=\frac{16\pi G}{c^4} \widetilde{\im}\left[\pn{\tau}\mn\right]
 - {4G\over c^4} \sum^{+\infty}_{l=0} {(-)^l\over l!} \hat{\partial}_L
 \left\{ \frac{\overline{{\cal A}\mn_L (t-r/c)-{\cal A}\mn_L
 (t+r/c)}}{2r} \right\}
\;, \end{equation} 
where the radiation-reaction function is composed of two terms~:

\begin{equation}
{\cal A}^{\mu\nu}_L(t) ={\cal
F}^{\mu\nu}_L(t) +{\cal R}^{\mu\nu}_L(t)
\;. \end{equation}
The first term is nothing but the exterior multipole moment given by
Eq. (\ref{64}), and one can check that it contains the standard
radiation-reaction effect at the 2.5PN order.  The ${\cal
R}^{\mu\nu}_L$-term is defined by Eq. (\ref{25}), or, rather, the
post-Newtonian expansion of it, i.e.

\begin{equation}\label{66}
{\cal R}\mn_L(t) = \pf\int d^3{\mathbf y}~\!|\widetilde{{\mathbf
y}}|^B {\hat y}_L \overline{\int_1^{+\infty}dz~\!\gamma_l(z)~\!{\cal
M}(\tau\mn)({\mathbf y},t-z |{\mathbf y}|/c)} \;. \end{equation} This
term is quite interesting~: it depends on the non-linearities of the
exterior field, described by the gravitational source term ${\cal
M}(\tau\mn)$ (or, more precisely, the non-stationary part of it
\cite{NN}), which are to be computed by means of the
multipolar-post-Minkowskian algorithm of Refs. \cite{BD86,B93} (see in
particular Section III.D in Ref. \cite{B93} for some detailed
computations of this term). Physically the function ${\cal
R}^{\mu\nu}_L$ contains the effect of wave tails in the radiation
reaction force which arises at the 4PN order \cite{BD88,B93,B97}. It
is not difficult [using notably the formula (\ref{gam1}) below] to
derive the more explicit expression for the contribution of ${\cal
R}^{\mu\nu}_L$ to the anti-symmetric wave in Eq. (\ref{65})~:

\begin{eqnarray}\label{66'}
&&\hat{\partial}_L \left\{ \frac{\overline{{\cal R}\mn_L (t-r/c)-{\cal
 R}\mn_L (t+r/c)}}{2r} \right\} =\nonumber\\
 &&\quad\sum_{i=0}^l\frac{(-)^l(l+i)!}{2^i i!(l-i)!}
 \sum_{k=0}^{+\infty}\frac{\Dn{-k}(\hat{x}_L)}{c^{2k+l-i}}\pf\int
 d^3{\mathbf y}~\!|\widetilde{{\mathbf
 y}}|^B\frac{\hat{y}_L}{\absy^{l+i+1}}\partial_t^{2k+l-i}{\cal
 M}(\tau\mn)({\mathbf y},t-\absy/c) \;. \end{eqnarray} When they are
 computed by post-Minkowskian approximations, the remaining integrals
 will typically yield, after integration over the angles, some
 ``hereditary-like'' contributions, depending on the whole integrated
 past of the matter source (see Ref. \cite{B93}).

It is tempting to speculate that the second term in Eq. (\ref{65}),
made of the anti-symmetric multipolar waves parametrized by the
functions ${\cal A}^{\mu\nu}_L(t)$, can be regarded as the
contribution, in a sense to be made more precise, of the radiation
reaction forces at work inside the post-Newtonian source. [Indeed we
have checked that these functions contain the known radiation-reaction
terms at the dominant 2.5PN order as well as the dominant contribution
of tails at the 4PN order.] We shall leave for future work the
systematic study of this term, as well as the possibility to answer
the latter speculation.

\section{Harmonic-coordinate condition}\label{V}

The latter solution for the post-Newtonian field,
Eqs. (\ref{65})-(\ref{66'}), has been obtained without imposing, in an
explicit way, the condition of harmonic coordinates (\ref{2}). Indeed,
we have assumed this condition to be true, and we simply matched
together the post-Newtonian and multipolar-post-Minkowskian
expansions, satisfying the relaxed Einstein field equations (\ref{1})
in their respective domains. We found that the matching determines
uniquely the expressions of the multipole moments $X\mn_L(t)$ and
radiation-reaction functions $A\mn_L(t)$ as some functionals of the
stress-energy pseudo-tensor $\tau\mn$. However, we never used the
harmonic-coordinate condition during the matching; it was not
necessary for the formal determination of the unknown parameters
($X\mn_L$, $A\mn_L$).  Therefore, it is quite important to check that
our post-Newtonian solution is divergenceless as a consequence of the
conservation of the pseudo-tensor $\tau\mn$ [see Eq. (\ref{2})], so
that we really grasp a solution of the full Einstein field equations.

We check the divergenceless of $\pn{h}\mn$ directly on
Eq. (\ref{44}). We apply the $\partial_{\mu}$ operator on each side of
the equality~:

\begin{equation}
\partial_{\mu}\pn{h}\mn=\frac{16\pi G}{c^4}
\partial_{\mu}\widetilde{\im}\left[\pn{\tau}\mn\right]
+\partial_{\mu}\bigg[\sum_{l,k=0}^{+\infty}\frac{1}{c^{2k}}
\overset{(2k)}{B_L\mn}(t)~\!\Dn{-k}({\hat
x}_L)\bigg]\label{divdiv}
\;. \end{equation}
We must transform the two terms on the right-hand-side 
in order to make explicit the fact that these two terms are exactly
the opposite. The first term, that is to say, the divergence of the
$\widetilde{\im}$ operator, is not obvious since, even if time
derivatives commute with $\widetilde{\im}$, spatial derivatives do
not,

\begin{eqnarray}
\partial_{\mu}\widetilde{\im}\left[\pn{\tau}\mn\right]
=\widetilde{\im}\left[\partial_0\pn{\tau}^{0\nu}\right]
+\partial_i\widetilde{\im}\left[\pn{\tau}^{i\nu}\right]
\;. \end{eqnarray} $\widetilde{\im}$ is a sum of
$\Dn{-k-1}\partial_t^{2k}$, and spatial derivatives do not commute
with $\Dn{-k-1}$ because of the $|\widetilde{{\mathbf y}}|^{B}$ factor
[see Eq. (\ref{37}) for the exact expression]. To see how to tackle
this problem, let us start with the spatial divergence of
$\Dn{-k-1}$. We assume that $\pn{\tau}({\mathbf x},t)$ is a function
of the ``post-Newtonian'' type, i.e. satisfies the requirements
(\ref{31})-(\ref{32}). The $\partial_i$ derivative, in
Eq. (\ref{div1}), applies first to the $k^{\mathrm{th}}$-Poisson's
kernel but after having noticed that the $x^i$-derivative of this
kernel was equal to minus the $y^i$-derivative of it, we can make an
integration by part and distribute the $y^i$-derivative on
$|\widetilde{{\mathbf y}}|^{B}$ and on $\pn{\tau}$ so that

\begin{eqnarray}
&&\partial_i\Dn{-k-1}(\pn{\tau}({\mathbf
x},t))=-\frac{1}{4\pi}\pf\int\dy\abty^{B}\partial_{x^i}\Big(\frac{|{\mathbf
x} -{\mathbf y}|^{2k-1}}{(2k)!}\Big)\pn{\tau}({\mathbf y},t)=\nonumber\\
&&-\frac{1}{4\pi}\pf\int\dy\abty^{B}\frac{|{\mathbf x}-{\mathbf
y}|^{2k-1}}{(2k)!}\partial_i\pn{\tau}({\mathbf y},t)
-\frac{1}{4\pi}\pf\int\dy\partial_i(\abty^{B})\frac{|{\mathbf x}-{\mathbf
y}|^{2k-1}}{(2k)!}\pn{\tau}({\mathbf y},t)
\label{div1}
\;. \end{eqnarray}
The first term in the last line of the previous equality is equal to
$\Dn{-k-1}(\partial_i\pn{\tau})$. Now, let us concentrate on the last
term in the same line. We can, of course, write
$\partial_i(\abty^{B})=B n_i\,\abty^{B-1}/r_0$. Moreover, since
$\pn{\tau}({\mathbf y},t)$ is regular at the origin ($\absy=0$), the
integral is always convergent on any neighbourhood of the
origin. Translating these two remarks in the last integral of
Eq. (\ref{div1}), and since we take the finite part when $B=0$, this
last integral is zero, because of the explicit factor $B$, when
ranging from $\absy=0$ up to some arbitrary finite value $\absy={\cal
R}$. So, after replacing $|{\mathbf x}-{\mathbf y}|^{2k-1}$ by
$(2k)!~\!\Dn{-k}(|{\mathbf x}-{\mathbf y}|^{-1})$ we are left, in
Eq. (\ref{div2}), with one integral ranging over $\absy > {\cal R}$,

\begin{eqnarray}
&&\pf\int\dy\partial_i(\abty^{B})\frac{|{\mathbf x}-{\mathbf
y}|^{2k-1}}{(2k)!}\pn{\tau}({\mathbf y},t)\nonumber\\
&&=\pf\int_{\absy > {\cal
R}}\!\!\!\!\!\dy\,\partial_i(\abty^{B})\pmk\Big[\Dn{-k}(|{\mathbf
x}-{\mathbf y}|^{-1})\Big]\,\pmk(\pn{\tau})({\mathbf y},t)\nonumber\\
&&=\sum_{n=0}^{k}\sum_{l\geq
0}\frac{(-)^l}{l!}\Dn{-k+n}(\xl{x})\pf\int_{\absy > {\cal
R}}\dy\,\partial_i
(\abty^{B})\Dn{-n}(\dl{\absy^{-1}})\,\pmk(\pn{\tau})({\mathbf
y},t)\label{div2} \;. \end{eqnarray} In the last line of the previous
equation, we expanded the $k^{\mathrm{th}}$-Poisson's kernel for
$\absy\gg\absx$ using Eq. (\ref{121}). This is possible thanks to the
fact that ${\cal R}$ is arbitrary and may be chosen such that ${\cal
R}\gg\absx$. We also note that $\pn{\tau}$ turned into
$\pmk(\pn{\tau})$ because $\pn{\tau}=\pmk(\pn{\tau})$ in the far
zone. In this way,

\begin{eqnarray}
&&\partial_i\Dn{-1-k}\left[\pn{\tau}({\mathbf
x},t)\right]=\Dn{-1-k}\left[\partial_i\pn{\tau}({\mathbf
x},t)\right]\nonumber\\ &&-\frac{1}{4\pi}\sum_{n=0}^{k}\sum_{l\geq
0}\frac{(-)^l}{l!}\Dn{-k+n}(\xl{x})\pf\int_{\absy > {\cal
R}}\dy\,\partial_i
(\abty^{B})\Dn{-n}(\dl{\absy^{-1}})\,\pmk(\pn{\tau})({\mathbf
y},t)\label{div5}
\;, \end{eqnarray}
and we notice that the commutation of the spatial derivative and the
generalized $k^{\mathrm{th}}$-Poisson integral depends only on the
behaviour of $\pn{\tau}({\mathbf x},t)$ at spatial infinity. This fact was
foreseeable since for a function $\pn{\tau}({\mathbf x},t)$ with compact
support the commutation would be trivial. Thanks to the general result
given by Eq. (\ref{div5}), in which we replace $\pn{\tau}$ by
$\pn{\tau}^{i\nu}$, we can determine the spatial divergence of
$\widetilde{\im}(\pn{\tau}^{i\nu})$. We can then get
$\partial_{\mu}\widetilde{\im}\left[\pn{\tau}\mn\right]$ that is the
sum of $\widetilde{\im}\left[\partial_{\mu}\pn{\tau}\mn\right]$ and a
non-trivial term. Since $\partial_{\mu}\pn{\tau}\mn=0$, the result for
the first term of Eq. (\ref{divdiv}) reduces to the non-trivial term,
that is to say~:

\begin{eqnarray}
&&\frac{16\pi
G}{c^4}\partial_{\mu}\widetilde{\im}\left[\pn{\tau}\mn\right]\nonumber\\
&&=-\frac{4G}{c^4}\sum_{k,l,n\geq
0}\frac{(-)^l}{l!}\Dn{-k}(\xl{x})\pf\int_{\absy > {\cal
R}}\!\!\!\!\dy\,\partial_i(\abty^{B})
\Dn{-n}(\dl{\absy^{-1}})\,\overset{(2k+2n)}{\pmk(\pn{\tau}^{i\nu})}({\mathbf
y},t)\label{divd6} \;. \end{eqnarray} Now, we want to prove that the
second term in the right-hand-side of Eq. (\ref{divdiv}) is exactly
the opposite. In Eq. (\ref{dmupm1}), we expand this last term in its
{\it 3+1} form so that we can treat separately terms with time
derivative and terms with spatial derivative,
 
\begin{eqnarray}
\sum_{n,l\ge
0}\frac{1}{c^{2n}}\overset{(2n)}{B_L^{i\nu}}(t)
\partial_i\big[\Dn{-n}(\xl{x})\big]+\sum_{n,l\ge
0}\frac{1}{c^{2n}}
\partial_0\overset{(2n)}{B_L^{0\nu}}(t)\Dn{-n}(\xl{x})\label{dmupm1}
\;. \end{eqnarray}
The first term of Eq. (\ref{dmupm1}), thanks to a STF formula, can be
written without the use of spatial derivative. The index $i$ coming
from this derivative is distributed on the multi-index $L$ in the way

\begin{eqnarray}\label{stf}
\sum_{n,l\geq
0}\frac{1}{c^{2n}}\overset{(2n)}{B_L^{i\nu}}(t)\partial_i\big[\Dn{-n}(\xl{x})\big]
=\sum_{n,l\geq
0}\frac{1}{c^{2n}}\Big\{\frac{1}{2l+3}\overset{(2n)}{B_L^{i\nu}}(t)\Dn{-n+1}({\hat
x}_{iL})+\,l\,\overset{(2n)} {B_{iL-1}^{i\nu}}(t)\Dn{-n}(&&{\hat
x}_{L-1})\Big\}\;.\nonumber\\ &&\label{div10}
\end{eqnarray}
In the second term of Eq. (\ref{dmupm1}), we express the function
$B^{0\nu}_L$ in terms of ${\cal F}^{0\nu}_L$ and ${\cal R}^{0\nu}_L$ [{\it cf}.
Eqs. (\ref{48}) and (\ref{62b})] because the time derivative,
$\partial_0$, will act on the integrand of these two time-varying
moments~:
  
\begin{eqnarray}
\Dn{-n}(\xl{x})\partial_0
B_L^{0\nu}(t)=\frac{4G(-)^l}{c^{5+2l}(2l+1)!! l!}\Dn{-n}(\xl{x})\Big\{\partial_0
\overset{(2l+1)}{{\cal
R}_L^{0\nu}}(t)+\partial_0\overset{(2l+1)}{{\cal
F}_L^{0\nu}}(t)\Big\}\label{dmupm3}
\;. \end{eqnarray}
First, we investigate the case of $\partial_0{\cal F}_L^{0\nu}$, using
the formula (\ref{53'}), where the time derivative acts on
$\pn{\tau}^{0\nu}$,

\begin{eqnarray}
\Dn{-n}(\xl{x})\partial_0{\cal
F}_L^{0\nu}(t)=\Dn{-n}(\xl{x})\sum_{k\geq
0}\frac{1}{c^{2k}}\pf\int\dy\,\abty^{B}\Dn{-k}(\xl{y})
\partial_0\overset{(2k)}{\pn{\tau}^{0\nu}}({\mathbf y},t)
\;. \end{eqnarray}
We can replace $\partial_0\pn{\tau}^{0\nu}$ by
$-\partial_i\pn{\tau}^{i\nu}$ thanks to the conservation equation of
the pseudo-tensor. After integrating by part we get

\begin{eqnarray}
&&\Dn{-n}(\xl{x})\sum_{k\geq
0}\frac{1}{c^{2k}}\pf\int\dy\,\partial_i(\abty^{B})\Dn{-k}(\xl{y})
\overset{(2k)}{\pn{\tau}^{i\nu}}({\mathbf
y},t)\nonumber\\ &&+\Dn{-n}(\xl{x})\sum_{n\geq
0}\frac{1}{c^{2k}}\pf\int\dy\,\abty^{B}\partial_i(\Dn{-k}(\xl{y}))
\overset{(2k)}{\pn{\tau}^{i\nu}}({\mathbf
y},t)\label{dmupm2}
\;. \end{eqnarray}
The same STF formula as used in Eq. (\ref{stf}) enables one to
transform the second term of Eq. (\ref{dmupm2}) so that, at the end,
we get the definitive result~:

\begin{eqnarray}
\Dn{-n}(\xl{x})\partial_0{\cal F}_L^{0\nu}(t) &=&
\Dn{-n}(\xl{x})\sum_{k\geq
0}\frac{1}{c^{2k}}\pf\int\dy\,\partial_i(\abty^{B})\Dn{-k}
(\xl{y})\overset{(2k)}{\pn{\tau}^{i\nu}}({\mathbf y},t)\nonumber\\ & &
+l\,\Dn{-n}({\hat x}_{iL-1}){\cal
F}_{L-1}^{i\nu}(t)+\frac{\Dn{-n}(\xl{x})}{c^2(2l+3)}\overset{(2)}
{{\cal F}_{iL}^{i\nu}}(t)\label{dumpm7}
\;. \end{eqnarray}

We can, now, investigate the case of the first term in
Eq. (\ref{dmupm3}), which is a little bit more complicated since it
involves a retarded integral,

\begin{eqnarray}
\Dn{-n}(\xl{x})\partial_0{\cal
R}_L^{0\nu}(t)=\Dn{-n}(\xl{x})\pf\int\dy\,\abty^{B}\xl{y}\int_{1}^{\infty}\,
\dz~\!  \gamma_l(z)\pmk(\partial_0\tau^{0\nu})({\mathbf y},t-z\absy/c)
\;, \end{eqnarray}
where the function $\gamma_l(z)$ is given by Eq. (\ref{26}) (for
simplicity's sake we do not write the overline indicating the
post-Newtonian expansion).  We do the replacement of
$\partial_0\tau^{0\nu}$ into $-\partial_i\tau^{i\nu}$. Before
integrating by part, we better have to notice that the partial
derivative $\partial_i$ acts on $\tau^{i\nu}$ which is then evaluated
at the event $({\mathbf y},t-z\absy/c)$; we must be careful about the
space dependence of the time variable $t-z\absy/c$. The last equation
then becomes

\begin{eqnarray}
&&-\Dn{-n}(\xl{x})\pf\int\dy\,
\abty^{B}\xl{y}\partial_i\left(\int_{1}^{\infty}\, \dz~\!\gamma_l(z)
\pmk(\tau^{i\nu})({\mathbf y},t-z\absy/c)\right)\nonumber\\
&&-\Dn{-n}(\xl{x})\pf\int\dy\, \abty^{B}\xl{y}n_i\int_{1}^{\infty}\,
\dz ~\!\frac{z}{c}~\!\gamma_l(z)\overset{(1)} {\pmk(\tau^{i\nu})}({\mathbf
y},t-z\absy/c)\label{dmupm4}
\;. \end{eqnarray}
In this way, the first term can be integrated by part
straigthforwardly, in terms of $\dy$ integration, showing up a
$\partial_i(\abty^{B})$ term and a $\partial_i(\xl{y})$ term. The
second term will also be integrated by part, in terms of $\dz$
integration, using the fact
$\frac{d}{dz}[\gamma_{l+1}(z)]=-(2l+3)z\gamma_l(z)$; so we have

\begin{eqnarray}
&&\Dn{-n}(\xl{x})\pf\int\dy\,
\partial_i(\abty^{B})~\!\xl{y}\int_{1}^{\infty}\, \dz~\!\gamma_l(z)
\pmk(\tau^{i\nu})({\mathbf y},t-z\absy/c)\nonumber\\
&&+\Dn{-n}(\xl{x})\pf\int\dy\,\abty^{B}\partial_i(\xl{y})\int_{1}^{\infty}\,
\dz~\!\gamma_l(z) \pmk(\tau^{i\nu})({\mathbf y},t-z\absy/c)\nonumber\\
&&+\frac{\Dn{-n}(\xl{x})}{c^2(2l+3)}\pf\int\dy\,\abty^{B}\xl{y}y_i\int_{1}^{\infty}\,
\dz~\!\gamma_{l+1}(z) \overset{(2)}{\pmk(\tau^{i\nu})}({\mathbf
y},t-z\absy/c)\label{dumpm5}
\;. \end{eqnarray}
The sum of these three terms can be transformed so that the function
${\cal R}_L^{i\nu}$ shows up. Since for any STF tensor
${\hat{T}}_L\partial_i(\xl{y})=l{\hat{T}}_{iL-1}{\hat y}_{L-1}$ and
${\hat{T}}_L\xl{y}y_i ={\hat{T}}_{L}{\hat
y}_{iL}+\frac{l}{2l+1}{\hat{T}}_{iL-1}{\hat y}_{L-1}\absy^2$, and keeping
in mind that all the multi-indices $L$ will have to be summed, we can
write

\begin{eqnarray}
&&\Dn{-n}(\xl{x})\pf\int\dy\,
\partial_i(\abty^{B})~\!\xl{y}\int_{1}^{\infty}\, \dz~\!\gamma_l(z)
\pmk(\tau^{i\nu})({\mathbf y},t-z\absy/c)\\ &&+l\,\Dn{-n}({\hat
x}_{iL-1})\pf\int\dy\,\abty^{B}{\hat y}_{L-1}\int_{1}^{\infty}\,
\dz~\!\gamma_l(z) \pmk(\tau^{i\nu})({\mathbf y},t-z\absy/c)\nonumber\\
&&+\frac{\Dn{-n}(\xl{x})}{c^2(2l+3)}\pf\int\dy\,\abty^{B}{\hat
y}_{iL}\int_{1}^{\infty}\, \dz~\!\gamma_{l+1}(z)
\overset{(2)}{\pmk(\tau^{i\nu})}({\mathbf y},t-z\absy/c)\nonumber\\
&&+\frac{l\Dn{-n}({\hat
x}_{iL-1})}{c^2(2l+1)(2l+3)} \pf\int\dy\,\abty^{B}\absy^2{\hat
y}_{L-1}\int_{1}^{\infty}\, \dz~\!\gamma_{l+1}(z)
\overset{(2)}{\pmk(\tau^{i\nu})}({\mathbf y},t-z\absy/c)\nonumber
\;. \end{eqnarray}
An interesting relation between $\gamma_l$-functions~:
$\frac{d^2}{dz^2}[\gamma_{l+1}(z)]=(2l+1)(2l+3)\big[\gamma_{l-1}(z)-\gamma_l(z)\big]$,
after integrating by part the last integral, in terms of $\dz$
integration, allows us to get the more explicit form :

\begin{eqnarray}
\Dn{-n}(\xl{x})\partial_0{\cal
R}_L^{0\nu}(t)&=&\Dn{-n}(\xl{x})\pf\int\dy\,
\partial_i(\abty^{B})~\!\xl{y}\int_{1}^{\infty}\,
\dz~\!\gamma_l(z)\pmk(\tau^{i\nu})({\mathbf y},t-z\absy/c)\nonumber\\ &&
+l\,\Dn{-n}({\hat x}_{iL-1}){\cal
R}_{L-1}^{i\nu}(t)+\frac{\Dn{-n}(\xl{x})}{c^2(2l+3)}
\overset{(2)}{{\cal R}_{iL}^{i\nu}}(t)\label{dumpm6}
\;. \end{eqnarray}
Summing up Eqs. (\ref{dumpm7}) and (\ref{dumpm6}) we obtain

\begin{eqnarray}
&&\sum_{n\ge
0}\partial_0\overset{(2n)}{B_L^{0\nu}}(t)\Dn{-n}(\xl{x})=\frac{4G}{c^4}\sum_{l,n\geq0}
\frac{2^l (-)^l}{(2l+1)!}\Dn{-n}(\xl{x})\bigg[\sum_{k\geq
0}\pf\int\dy\,\partial_i(\abty^{B})\Dn{-n}(\xl{y})
\frac{\overset{(2n+2k+2l+1)}{\pn{\tau}^{i\nu}({\mathbf
y},t)}}{c^{2n+2k+2l+1}}\nonumber\\ &&\qquad\quad +\pf\int\dy\,
\partial_i(\abty^{B})\xl{y}\int_{1}^{\infty}\,
\dz~\!\gamma_l(z)\frac{\overset{(2n+2l+1)}{\pmk(\tau^{i\nu})}}{c^{2n+2l+1}}({\mathbf
y},t-z\absy/c) \bigg]\nonumber\\ &&\qquad\quad -\sum_{n,l\geq
0}\frac{1}{c^{2n}}\Big\{\frac{1}{2l+3}\overset{(2n)}{B_L^{i\nu}}(t)\Dn{-n+1}({\hat
x}_{iL})+\,l\,\overset{(2n)} {B_{iL-1}^{i\nu}}(t)\Dn{-n}({\hat
x}_{L-1})\Big\}\label{divd18} \;. \end{eqnarray} The last line cancels
out the terms coming from Eq. (\ref{div10}).

We can therefore write down the result for the divergence of
$\pn{h}\mn$ which, at this stage, depends only on terms with integrals
of $\partial_i(\abty^{B})$ and having the spatial structure given by
$\Dn{-n}(\xl{x})$. After summing Eqs. (\ref{divd6}), (\ref{div10}) and
(\ref{divd18}) we get

\begin{eqnarray}
&&\partial_{\mu}\pn{h}\mn=\label{dumpm9}\\ &&-\frac{4G}{c^4}\sum_{n,l,k\geq
0}\frac{(-)^l}{l!}\frac{\Dn{-n}(\xl{x})}{c^{2k+2n}} \pf\int_{\absy >
{\cal
R}}\dy\,\partial_i(\abty^{B})\Dn{-k}(\dl{\absy^{-1}})\,
\overset{(2n+2k)}{\pmk(\pn{\tau}^{i\nu})}
({\mathbf y},t)\nonumber\\ &&+\frac{4G}{c^4}\sum_{n,l,k\geq 0}\frac{2^l
(-)^l}{(2l+1)!}\frac{\Dn{-n}(\xl{x})}{c^{2n+2k+2l+1}}\pf\int_{\absy >
{\cal R}}\dy\,\partial_i(\abty^{B})
\Dn{-k}(\xl{y})\overset{(2n+2k+2l+1)}{\pmk(\pn{\tau}^{i\nu})({\mathbf
y},t)}\nonumber\\ &&+\frac{4G}{c^4}\sum_{n,l\geq 0}\frac{2^l
(-)^l}{(2l+1)!}\frac{\Dn{-n}(\xl{x})}{c^{2n+2l+1}}\pf\int\dy\,
\partial_i(\abty^{B}) ~\!\xl{y}\int_{1}^{\infty}\,
\dz~\!\gamma_l(z)\overset{(2n+2l+1)}{\pmk(\tau^{i\nu})}({\mathbf
y},t-z\absy/c)\nonumber
\;. \end{eqnarray}
In the second term we have used the fact that the integral depends
only on the values for which $\absy > {\cal R}$ to write
$\pn{\tau}^{i\nu}=\pmk(\pn{\tau}^{i\nu})$ on that domain.  The last
term of Eq. (\ref{dumpm9}) depends on a retarded integral of the
multipolar post-Minkowskian expansion $\pmk(\tau^{i\nu})$. By
integrating by part the integral over $z$ one can transform this last
term into
 
\begin{eqnarray}
\frac{4G}{c^4}\sum_{l,n\geq 0}&&\frac{2^l
(-)^l}{(2l+1)!}\Dn{-n}(\xl{x})\nonumber\\ &&\times\sum_{p\geq
0}\frac{1}{c^{2n-p+l}}\pf\int_{\absy < {\cal R}}\dy\,
\partial_i(\abty^{B})\xl{y}\absy^{-p-l-1}\,\overset{(p+l)}
{\gamma_l}\!(1)\,\overset{(2n-p+l)}{\pmk(\tau^{i\nu})}({\mathbf y},t
-\absy/c)\label{dumpm10} \;. \end{eqnarray} The superscript $(p+l)$ on
the $\gamma_l$-function refers to the $z$-differentiation. It is
straightforward to show, using the fact that
$\overset{(l)}{\gamma_l}(z)=(-)^{l+1}(2l+1)!!P_l(z)$ is directly
related to the Legendre polynomial, that

\begin{equation}\label{gam1}
\overset{(p+l)}{\gamma_l}\!(1) = (-)^{l+1}\frac{(2l+1)!!(l+p)!}{2^p p!(l-p)!}
\;. \end{equation}
Since $\pmk(\tau^{i\nu})$ is singular at the origin (but regular at
infinity), and because of the explicit factor $B$ brought about by the
derivative $\partial_i(\abty^{B})$, the integral in
Eq. (\ref{dumpm10}) ranges over $\absy < {\cal R}$ (and even $\absy <
\epsilon$ where $\epsilon$ is an arbitrary small number).  We can then
expand $\pmk(\tau^{i\nu})({\mathbf y},t-\absy/c)$ when $c\rightarrow
+\infty$. Furthermore we can change the integration over $\absy <
{\cal R}$ into an integration over $\absy > {\cal R}$ by simply
changing the sign in front of the integral.  Indeed, this comes from a
technical lemma, which plays an important role in
Refs. \cite{B95,B98mult}; see before Eq. (\ref{116}) in Appendix
\ref{C}, and the proof given in Ref. \cite{N6}.  Thus,

\begin{eqnarray}
&&-\frac{4G}{c^4}\sum_{n,l,k\geq 0}\frac{2^l
(-)^l}{(2l+1)!}\Dn{-n}(\xl{x})\nonumber\\ &&\quad\times\sum_{p\geq
0}\frac{1}{c^{2n-p+l+k}}\pf\int_{\absy > {\cal R}}\dy\,
\partial_i(\abty^{B})\xl{n}\frac{(-)^k}{k!}
\absy^{k-p-1}\,\overset{(p+l)}{\gamma_l}\!(1)\,
\overset{(2n-p+l+k)}{\pmk(\tau^{i\nu})({\mathbf
y},t)}
\;. \end{eqnarray}
By changing the label $k$ into $2k+p-l$ and $2k+1+p-l$, in order to
cover odd and even numbers, we are able to write the previous
expression in terms of some sums of real numbers indexed by $p$, i.e.

\begin{eqnarray}
&&-\frac{4G}{c^4}\sum_{n,l,k\geq 0}\frac{2^l
(-)^l}{(2l+1)!}\Dn{-n}(\xl{x})\bigg\{\sum_{p=0}^{l}\frac{(-1)^{p+l}
\overset{(p+l)}{\gamma_l}\!(1)}{(p-l+2k)!}\bigg\}\nonumber\\
&&\quad\qquad\qquad\times\pf\int_{\absy > {\cal R}}\dy\,
\partial_i(\abty^{B})\xl{n}\absy^{2k-l-1}\overset{(2n+2k)}{\pmk(\tau^{i\nu})}({\mathbf
y},t)\nonumber\\ &&+\frac{4G}{c^4}\sum_{n,l,k\geq 0}\frac{2^l
(-)^l}{(2l+1)!}\Dn{-n}(\xl{x})
\bigg\{\sum_{p=0}^{l}\frac{(-1)^{p+l}\overset{(p+l)}
{\gamma_l}\!(1)}{(p-l+2k+1)!}\bigg\}\nonumber\\
&&\quad\qquad\qquad\times\pf\int_{\absy > {\cal R}}\dy\,
\partial_i(\abty^{B})\xl{n}\absy^{2k-l}\overset{(2n+2k+1)}{\pmk(\tau^{i\nu})}({\mathbf
y},t)\label{div20}
\;. \end{eqnarray}
The sums in curly brackets are found to be explicit expressions
depending on $k$ and $l$ and some factorial combinations~:

\begin{eqnarray}
\forall k\geq
l+1,~~~~\sum_{p=0}^{l}\frac{(-1)^{p+l}
\overset{(p+l)}{\gamma_l}\!(1)}{(p-l+2k)!}&=&-\frac{(2l+1)!!}{2^k
k! (2k-2l-1)!!}\label{apG1}\;,\\ \forall k\geq
l,~~~~\sum_{p=0}^{l}\frac{(-1)^{p+l}\overset{(p+l)}
{\gamma_l}\!(1)}{(p-l+2k+1)!}&=&-\frac{(2l+1)!!}{2^{k-l}
(k-l)! (2k+1)!!}\label{apG2}\;,\\ \forall k\leq
l,~~~~\sum_{p=0}^{l}\frac{(-1)^{p+l}\overset{(p+l)}
{\gamma_l}\!(1)}{(p-l+2k)!}&=&(-1)^{k+l+1}\frac{(2l+1)!!(2l-2k-1)!!}{2^k
k!}\label{apG3}\;,\\ \forall k\leq
l-1,~~~~\sum_{p=0}^{l}\frac{(-1)^{p+l}\overset{(p+l)}
{\gamma_l}\!(1)}{(p-l+2k+1)!}&=&0\label{apG4}
\;. \end{eqnarray}
Thanks to these formulas, one can transform Eq. (\ref{div20}) into

\begin{eqnarray}
&&\frac{4G}{c^4}\sum_{n,l\geq 0}\sum_{k\geq
l+1}\frac{(-)^l}{l!}\frac{1}{2^k k!(2k-2l-1)!!}\Dn{-n}(\xl{x})\nonumber\\
&&\quad\qquad\qquad\times\pf\int_{\absy > {\cal R}}\dy\,
\partial_i(\abty^{B})~\!\xl{n}\absy^{2k-l-1}\overset{(2n+2k)}{\pmk(\tau^{i\nu})}({\mathbf
y},t) \nonumber\\ &&+\frac{4G}{c^4}\sum_{n,l\geq 0}\sum_{k\leq
l}\frac{(-)^k(2l-2k-1)!!}{l!2^{k}k!}\Dn{-n}(\xl{x})\nonumber\\
&&\quad\qquad\qquad\times\pf\int_{\absy > {\cal R}}\dy\,
\partial_i(\abty^{B})~\!\xl{n}\absy^{2k-l-1}\overset{(2n+2k)}{\pmk(\tau^{i\nu})}({\mathbf
y},t) \nonumber\\ &&-\frac{4G}{c^4}\sum_{n,l\geq 0}\sum_{k\geq
l}\frac{(-)^l}{l!}\frac{1}{2^{k-l}(2k+1)!!(k-l)!}\Dn{-n}(\xl{x})\nonumber\\
&&\quad\qquad\qquad\times\pf\int_{\absy > {\cal R}} \dy\,
\partial_i(\abty^{B})~\!\xl{n}
\absy^{2k-l}\overset{(2n+2k+1)}{\pmk(\tau^{i\nu})}({\mathbf y},t)
\label{dumpm13}
\;. \end{eqnarray}
In the latter expression we can recognize

\begin{eqnarray}
\forall k\geq
l,~~~~\Dn{-k}(\dl{\absy^{-1}})&=&\frac{1}{(2k-2l-1)!!2^k
k!}~\!\xl{n}\absy^{2k-l-1}\label{apD2}\;,\\ \forall k\leq
l,~~~~\Dn{-k}(\dl{\absy^{-1}})&=&\frac{(-)^{k+l} (2l-2k-1)!!}{2^k k!
}~\!\xl{n}\absy^{2k-l-1}\label{apD3}\;,\\
\Dn{-k}(\xl{y})&=&\frac{(2l+1)!!}{2^k k!(2k+2l+1)!!}~\!\xl{y}
\absy^{2k}\label{apD1}
\;, \end{eqnarray}
so that we obtain

\begin{eqnarray}
&&\frac{4G}{c^4}\sum_{l,n\geq 0}\frac{2^l
(-)^l}{(2l+1)!}\Dn{-n}(\xl{x})\pf\int\dy\,
\partial_i(\abty^{B})\xl{y}
\int_{1}^{\infty}\dz~\!\gamma_l(z)\overset{(2n+2l+1)}{\pmk(\tau^{i\nu})}({\mathbf
y},t-z\absy/c) \nonumber\\&&\quad = \frac{4G}{c^4}\sum_{n,l,k\geq
0}\frac{(-)^l}{l!}\Dn{-n}(\xl{x})\pf\int_{\absy > {\cal
R}}\dy\,\partial_i(\abty^{B})\Dn{-k}(\dl{\absy^{-1}})
\,\overset{(2n+2k)}{\pmk(\pn{\tau}^{i\nu})})({\mathbf y},t)\nonumber\\
&&\quad-\frac{4G}{c^4}\sum_{n,l,k\geq 0}\frac{2^l
(-)^l}{(2l+1)!}\Dn{-n}(\xl{x})\pf\int_{\absy > {\cal R}}\dy\,
\partial_i(\abty^{B})\Dn{-k}(\xl{y})\overset{(2n+2k+2l+1)}{\pmk(\pn{\tau}^{i\nu})({\mathbf
y},t)}\label{div30}
\;. \end{eqnarray}
After replacing Eq. (\ref{div30}) in Eq. (\ref{dumpm9}), at long last we find

\begin{equation}
\partial_{\mu}\pn{h}\mn=0
\;. \end{equation}
In this way, we have checked that the post-Newtonian metric, found by
matching as a definite functional of the stress-energy pseudo-tensor
$\tau\mn$, satisfies the harmonic-coordinate condition as a
consequence of the conservation of this pseudo-tensor.

\appendix

\section{Near-zone expansion of the retarded integral}\label{A}

This appendix, provided here for completeness, is an extended, and
also somewhat simplified, version of the derivation given in Appendix
A of Ref. \cite{B93}.  We are interested in source functions, say
${\cal M}(\tau)({\mathbf x},t)$, having the form of an exterior
multipole-moment decomposition, valid outside the compact-support
domain of the source. We employ the same notation as in Section
\ref{IIA} (except that we do not write the space-time indices)~:
$\tau$ denotes the pseudo-tensor of the source; notably we have ${\cal
M}(\tau)=\frac{c^4}{16\pi G}{\cal M}(\Lambda)$ where $\Lambda$ is the
gravitational source term. The two basic properties of the function
${\cal M}(\tau)({\mathbf x},t)$ are that it is smooth on $\Real^4$
deprived from the spatial origin $r=0$~:

\begin{equation}\label{67}
{\cal M}(\tau)({\mathbf x},t)\in C^\infty (\Real_*^3\times\Real)
\;, \end{equation}
and that it admits a near-zone expansion, when $r\to 0$ (with $t=$
const), having the appropriate structure [{\it cf} Eq. (\ref{17})]~:
i.e., $\forall N\in\N$,

\begin{equation}\label{68}
{\cal M}(\tau)({\mathbf x},t)=~ \sum {\hat n}_L r^a (\ln r)^p
G_{L,a,p}(t) + {\cal O}(r^N) 
\;, \end{equation}
where $a\in\Z$ with $a\leq N$ and $p\in\N$.  Like in Section
\ref{IIB} we denote with an overline the formal (infinite) near-zone
expansion,

\begin{equation}\label{69}
\overline{{\cal M}(\tau)({\mathbf x},t)}=~ \sum {\hat n}_L r^a (\ln
r)^p G_{L,a,p}(t) \;. \end{equation} It is very important to make the
distinction between ${\cal M}(\tau)$ and its formal near-zone
expansion $\overline{{\cal M}(\tau)}$.  Here we shall investigate the
retarded integral of the product $r^B{\cal M}(\tau)({\mathbf x},t)$,
where $B\in\C$, by means of analytic continuation (we pose $r_0=1$ in
this Appendix). For this task we assume at first that the real part of
$B$ is large enough so as to ``kill'' the divergencies, when $r\to 0$,
of the expansion (\ref{68}), so that the retarded integral is
initially well-defined.  Therefore, rigorously speaking, we are
allowed to do this only if there exists a finite maximal divergency,
i.e. some $a_\mathrm{min}\leq a$ in Eq. (\ref{68}) with finite
$a_\mathrm{min}\in\Z$. We have seen in Section \ref{IIA} that such
maximal divergency exists at any given post-Minkowskian order $m$, but
no longer exists for the full post-Minkowskian series because
$a_\mathrm{min}(m)\to -\infty$ when $m\to +\infty$. The consequence
is that the analytic continuation is in principle justified only at a
given finite post-Minkowskian order. But, as explained in Section
\ref{IIA}, we sum up systematically all the post-Minkowskian
results. In this way we are entitled to proceed as we do below; simply
we have to remember that the end result will be {\it a priori} true
only in a sense of formal post-Minkowskian expansions.

We decompose the source term into multipoles according to

\begin{equation}\label{70}
{\cal M}(\tau)({\mathbf x},t) = \sum_{l=0}^{+\infty} {\hat n}_L \sigma_L(r,t)
\;, \end{equation}
where the $\sigma_L$'s are STF functions in $L=i_1\cdots i_l$. The
inverse formula is

\begin{equation}\label{71}
\sigma_L(r,t) = \frac{(2l+1)!!}{l!}\int\frac{d\Omega}{4\pi} ~\!{\hat
n}_L {\cal M}(\tau)({\mathbf x},t)
\;, \end{equation}
where $d\Omega$ is the solid-angle element around the unit vector
$n_i=n^i=x^i/r$. Then the expression of the retarded integral, in a
sense of analytic continuation in $B$, is given by the following
explicit formula, obtained in Ref. \cite{BD86} [see Eqs. (6.3)-(6.5)
there]~:

\begin{equation}\label{72}
\Box^{-1}_\mathrm{Ret}\left[r^B{\cal M}(\tau)({\mathbf x},t)\right] =
\sum_{l=0}^{+\infty} \int_{-\infty}^{t-r}ds~\hat{\partial}_L
\left\{\frac{R^B_L\left(\frac{t-r-s}{2},s\right)-R^B_L
\left(\frac{t+r-s}{2},s\right)}{r}\right\}
 \end{equation}
(we pose $c=1$ and $r_0=1$ in this Appendix), where the function $R^B_L(\rho,s)$
reads

\begin{equation}\label{73}
R^B_L(\rho,s) = \rho^l\int_0^\rho
dx~\frac{(\rho-x)^l}{l!}\left(\frac{2}{x}\right)^{l-1} x^B\sigma_L(x,x+s)
\;. \end{equation}
Following the same procedure as in Eqs. (A6)-(A7) in Ref. \cite{B93}
we are allowed to re-write the expression (\ref{72}) into the
alternative form

\begin{eqnarray}\label{74}
\Box^{-1}_\mathrm{Ret}\left[r^B{\cal M}(\tau)({\mathbf x},t)\right] &=&
\sum_{l=0}^{+\infty} \int_{-r}^{r}du~\hat{\partial}_L
\left\{\frac{1}{r}R^B_L\left(\frac{u+r}{2},t-u\right)\right\}\nonumber\\
&-&\frac{1}{4\pi}\sum_{l=0}^{+\infty}\frac{(-)^l}{l!}\hat{\partial}_L\left[\frac{{\cal
R}^B_L(t-r)-{\cal R}^B_L(t+r)}{2r}\right]
\;. \end{eqnarray}
The ``anti-symmetric'' wave is parametrized by ${\cal R}^B_L(t)$ which
is related to the function $R^B_L(\rho,s)$ by

\begin{equation}\label{75}
{\cal R}^B_L(t) = 8\pi (-)^{l+1}l!\int_{-\infty}^t
ds~R^B_L\left(\frac{t-s}{2},s\right)
\;. \end{equation}
Inserting Eq. (\ref{73}), and performing some change of variables, we
obtain

\begin{equation}\label{76}
{\cal R}^B_L(t) = \frac{4\pi l!}{(2l+1)!!} ~\!\int_0^{+\infty}
dx~\!x^{B+l+2}\int_1^{+\infty}dz~\!\gamma_l(z)~\!\sigma_L(x,t-z x)
\;, \end{equation}
and, using the relation (\ref{71}), and considering the variable $x$ as the
norm of ${\mathbf x}\in\Real^3$, we further get

\begin{equation}\label{77}
{\cal R}^B_L(t) = \int d^3{\mathbf x}~\!|{\mathbf x}|^B {\hat x}_L
\int_1^{+\infty}dz~\!\gamma_l(z)~\!{\cal M}(\tau)({\mathbf x},t-z
|{\mathbf x}|) \;. \end{equation} In these expressions the function
$\gamma_l(z)$ is defined by

\begin{equation}\label{78}
\gamma_l(z)=(-)^{l+1}\frac{(2l+1)!!}{2^l l!}(z^2-1)^l
\;, \end{equation}
where the particular $l$-dependent factor has been chosen in such a
way that the integral is normalized to one in the following sense (see
Ref. \cite{B93}). Considering first that $l$ is a complex number such
that $-1<\Re(l)<-1/2$ we can compute the integral of $\gamma_l(z)$ by
means of the Euler $\Gamma$-function, with the result

\begin{equation}\label{79}
\int_1^{+\infty} dz~\!\gamma_l(z) =
2(-)^{l+1}\frac{\Gamma(2l+2)\Gamma(-2l-1)}{\Gamma(l+1)\Gamma(-l)}
\;. \end{equation} The right-hand-side of this equation can be
analytically continued to all values $l\in\C$ except half-integer
values, and is found to be equal to one when $l$ is an integer~:

\begin{equation}\label{80}
\int_1^{+\infty} dz~\!\gamma_l(z) = 1\qquad\hbox{($l\in\N$)}
\;. \end{equation}
Next, let us treat the first term in the right-hand-side of
Eq. (\ref{74}), say

\begin{equation}\label{81}
J^B({\mathbf x},t) \equiv \sum_{l=0}^{+\infty}
\int_{-r}^{r}du~\hat{\partial}_L
\left\{\frac{1}{r}R^B_L\left(\frac{u+r}{2},t-u\right)\right\}
\;. \end{equation}
This term is a particular solution of the d'Alembertian equation $\Box
J^B = r^B{\cal M}(\tau)$ [since the second term in Eq. (\ref{74}) is a
source-free solution].  We shall prove that the (formal) near-zone
expansion of that term, i.e. $\overline{J^B({\mathbf x},t)}$, is given by
the integral of the ``instantaneous'' potentials acting on the
near-zone expansion of the source term, i.e. $r^B\overline{{\cal
M}(\tau)({\mathbf x},t)}$. For any of the terms composing the multipolar
source $r^B\overline{{\cal M}(\tau)}$ [see Eq. (\ref{69})], we first
define

\begin{equation}\label{82}
\Delta^{-1}\left[{\hat n}_L r^{B+a} (\ln r)^p G_{L,a,p}(t)\right] =
\left(\frac{d}{dB}\right)^p\left[\frac{{\hat n}_L r^{B+a+2}
G_{L,a,p}(t)}{(B+a+2-l)(B+a+3+l)}\right] \end{equation} (this being
justified by the fact that one gets an identity by applying $\Delta$
on both sides).  Clearly the previous formula can be iterated and so
we can define the operator $\Delta^{-k-1}\equiv (\Delta^{-1})^{k+1}$,
applied on each separate terms in Eq. (\ref{69}) and therefore on the
complete series $r^B\overline{{\cal M}(\tau)({\mathbf x},t)}$. From
this we obtain the instantaneous-potentials operator, as the formal
expansion series
 
\begin{equation}\label{83}
\im\left[r^B\overline{{\cal M}(\tau)({\mathbf x},t)}\right] =
\sum_{k=0}^{+\infty}\left(\frac{\partial} {c\partial
t}\right)^{2k}\Delta^{-k-1}\left[r^B\overline{{\cal M}(\tau)({\mathbf
x},t)}\right]
\;. \end{equation}
Notice that this operator $\im$ contains only some even powers of
$1/c$. An important point for our purpose is that
$\im\big[r^B\overline{{\cal M}(\tau)}\big]$ is {\it proportional} to
the regularization factor $r^B$; and it evidently satisfies
$\Box\big(\im\big[r^B\overline{{\cal M}(\tau)}\big]\big)=
r^B\overline{{\cal M}(\tau)}$. On the other hand, we have also the
equation $\Box\overline{J^B} = r^B\overline{{\cal M}(\tau)}$, which
comes from applying the overline operation onto $\Box J^B = r^B{\cal
M}(\tau)$. This shows that $r^B\overline{{\cal M}(\tau)}$ and
$\overline{J^B}$ must differ by a solution of the homogeneous
equation, hence there should exist some functions $C_L^B(t)$ and
$D_L^B(t)$ such that

\begin{equation}\label{84}
\overline{J^B({\mathbf x},t)} = \im\left[r^B\overline{{\cal
M}(\tau)({\mathbf x},t)}\right]+\sum_{l=0}^{+\infty}\hat{\partial}_L
\left\{\frac{\overline{C_L^B(t-r)+D_L^B(t+r)}}{r}\right\}
\;. \end{equation} Note that the dependence on $B$ of the second term
is ``hidden'' inside the functions $C_L^B$ and $D_L^B$.  Let us now
prove that in fact the latter functions must be zero. This is a simple
consequence of the expression (\ref{73}) for the function
$R^B_L(\rho,s)$, from which we deduce that the expansion when $\rho\to
0$ of this function is proportional to $\rho^B$; in fact, it has the
structure $R^B_L(\rho,s)\sim \sum \rho^{B+b}(\ln\rho)^q$, when
$\rho\to 0$.  From this knowledge, we easily find that the near-zone
expansion of $\overline{J^B}$ is proportional to the factor
$r^B$. Since, as we have remarked, this is also the case of the first
term in Eq. (\ref{84}), $\im\big[r^B\overline{{\cal M}(\tau)}\big]$,
and since it is impossible that (the near-zone expansion of) the
second term in Eq. (\ref{84}) be itself proportional to $r^B$ -- the
$B$'s affect only the functions $C_L^B$ and $D_L^B$ but not the
structure of the near-zone expansion -- we conclude that $C_L^B$ and
$D_L^B$ are identically zero. Hence we have proved

\begin{equation}\label{85}
\overline{J^B({\mathbf x},t)} = \im\left[r^B\overline{{\cal M}(\tau)({\mathbf
x},t)}\right]
\;. \end{equation}
It suffices now to apply the overline operation (i.e., to take the
near-zone expansion) onto Eq. (\ref{74}) to get our final result,

\begin{equation}\label{86}
\overline{\Box^{-1}_\mathrm{Ret}\left[r^B{\cal M}(\tau)({\mathbf
x},t)\right]} = \im\left[r^B\overline{{\cal M}(\tau)({\mathbf
x},t)}\right]
-\frac{1}{4\pi}\sum_{l=0}^{+\infty}\frac{(-)^l}{l!}
\hat{\partial}_L\left[\frac{\overline{{\cal
R}^B_L(t-r)-{\cal R}^B_L(t+r)}}{2r}\right]
\;, \end{equation}
where we recall that the function ${\cal R}^B_L(t)$ has been given by
Eq. (\ref{77}). [The formula used Section \ref{IIB} results from
applying the finite part operation $\pf$.]

\section{The generalized Poisson operator}\label{B}

In Appendix \ref{A} we have been interested in source functions of the
multipolar type ${\cal M}(\tau)({\mathbf x},t)$, which are smooth in ${\mathbb
R}_*^3\times\Real$ and possess a {\it near-zone} expansion of the
type (\ref{69}). In the present Appendix \ref{B} we consider some
source functions of the post-Newtonian type $\pn{\tau}({\mathbf
x},t)$. These are supposed to be smooth all over $\Real^4$,

\begin{equation}\label{87}
\pn{\tau}({\mathbf x},t) ~\in~ C^\infty(\Real^4)
\;, \end{equation}
and to admit a {\it far-zone} expansion with structure ($\forall N\in\N$)

\begin{equation}\label{88}
\pn{\tau}({\mathbf x},t) =~ \sum {\hat n}_L r^a (\ln r)^p
G_{L,a,p}(t) + S_N({\mathbf x},t)
\;, \end{equation}
where $a\in\Z$, with $-N\leq a$, and $p\in\N$.  The
remainder-term is $S_N({\mathbf x},t) = {\cal O}\left(1/r^N\right)$ when
$r\to +\infty$ with $t=$ const.

Let us consider some $B\in\C$, and a radius ${\cal R}\in\Real$
with ${\cal R}>0$.  We define two integrals, corresponding to a split
of the Poisson integral between ``near-zone'' and ``far-zone''
contributions, separated by the radius ${\cal R}$~:

\begin{eqnarray}\label{89}
I_<^B({\mathbf x},t) &=& -\frac{1}{4\pi}\int_{|{\mathbf y}|<{\cal
R}}\frac{d^3{\mathbf y}}{|{\mathbf x}-{\mathbf y}|} ~\!|\widetilde{{\mathbf
y}}|^B\pn{\tau}({\mathbf y},t)\label{89a}\;,\\ I_>^B({\mathbf x},t) &=&
-\frac{1}{4\pi}\int_{|{\mathbf y}|>{\cal R}}\frac{d^3{\mathbf y}}{|{\mathbf
x}-{\mathbf y}|} ~\!|\widetilde{{\mathbf y}}|^B\pn{\tau}({\mathbf y},t)\label{89b}
\;. \end{eqnarray}
The $B$-dependent regularization factor is $|\widetilde{{\mathbf
y}}|^B\equiv (|{\mathbf y}|/r_0)^B$. It is easily checked that the
near-zone integral $I_<^B({\mathbf x},t)$ is well-defined (convergent)
when $\Re(B)>-3$ and that the far-zone one $I_>^B({\mathbf x},t)$ is
well-defined when $\Re(B)<-a_\mathrm{max}-2$, where $a_\mathrm{max}$ is the
maximal power of $r$ in the expansion (\ref{88}). So we have to assume
at this stage the existence of some maximal divergency corresponding
to some power $a_\mathrm{max}$. Strictly speaking, our present
investigation is thus valid only at some finite post-Newtonian
order. But, {\it in fine}, we sum up the results, and we consider the
complete post-Newtonian series to hold true in a formal sense.

We want first to check that the integrals (\ref{89a}) and (\ref{89b})
can be analytically continued down to a neighbourhood of $B=0$ (except
at the value $B=0$ itself), let say in the open domain ${\cal
B}_\epsilon$ defined by $0<|B|<\epsilon$ (where $\epsilon < 1$).
There is no problem with the near-zone integral $I_<^B({\mathbf x},t)$
which is clearly convergent all over ${\cal B}_\epsilon$ and even at
the value $B=0$. Concerning the far-zone integral $I_>^B({\mathbf
x},t)$ we replace the function $\pn{\tau}$ inside the integrand by its
far-zone expansion (\ref{88})~:

\begin{equation}\label{90}
I_>^B({\mathbf x},t) = -\frac{1}{4\pi}\int_{|{\mathbf y}|>{\cal
R}}\frac{d^3{\mathbf y}~\!|\widetilde{{\mathbf y}}|^B}{|{\mathbf x}-{\mathbf
y}|}\left\{\sum {\hat n}_L({\mathbf y}) |{\mathbf y}|^a (\ln |{\mathbf y}|)^p
G_{L,a,p}(t) + S_N({\mathbf y},t)\right\}
\;. \end{equation}
When $N$ is large enough the contribution due to the remainder $S_N$
is convergent all over ${\cal B}_\epsilon$ and at $B=0$, with
evidently the value at $B=0$ given by

\begin{equation}\label{91}
\int_{|{\mathbf y}|>{\cal R}}\frac{d^3{\mathbf y}~\!|\widetilde{{\mathbf
y}}|^B}{|{\mathbf x}-{\mathbf y}|} S_N({\mathbf y},t) = \int_{|{\mathbf y}|>{\cal
R}}\frac{d^3{\mathbf y}}{|{\mathbf x}-{\mathbf y}|} S_N({\mathbf y},t) + {\cal O}(B)
\;. \end{equation}
Thus we need only to deal with the other contributions, which consist
of a finite sum of terms, say

\begin{equation}\label{92}
\sum\int_{|{\mathbf y}|>{\cal R}}\frac{d^3{\mathbf y}~\!|\widetilde{{\mathbf
y}}|^B}{|{\mathbf x}-{\mathbf y}|}~\!{\hat n}_L({\mathbf y}) |{\mathbf y}|^a (\ln
|{\mathbf y}|)^p
\;. \end{equation}
Let us suppose that the field point ${\mathbf x}$ lies inside the far-zone
domain, i.e. ${\cal R}<|{\mathbf x}|$. We distinguish the two cases where
$|{\mathbf y}|<|{\mathbf x}|$ and $|{\mathbf x}|<|{\mathbf y}|$. For each of these two
cases we substitute into the integrals the appropriate multipolar
expansion of the factor $\frac{1}{|{\mathbf x}-{\mathbf y}|}$, for instance
$\frac{1}{|{\mathbf x}-{\mathbf
y}|}=\sum_{l=0}^{+\infty}\frac{(-)^l}{l!}y^L\hat{\partial}_L\frac{1}{|{\mathbf
x}|}$ when $|{\mathbf y}|<|{\mathbf x}|$.  This leads, after performing the
integration over the angles, to some series of radial integrals having
the structure (ignoring some unimportant factors)

\begin{equation}\label{93}
\sum\frac{{\hat x}_L}{|{\mathbf x}|^{2l+1}}\int_{{\cal R}}^{|{\mathbf
x}|}d|{\mathbf y}|~\!|{\mathbf y}|^{B+a+l+2} (\ln |{\mathbf y}|)^p + \sum {\hat
x}_L\int_{|{\mathbf x}|}^{+\infty}d|{\mathbf y}|~\!|{\mathbf y}|^{B+a-l+1} (\ln
|{\mathbf y}|)^p
\;. \end{equation}
When $|{\mathbf x}|<{\cal R}$ the reasoning is the same but simply one
ignores the first term in Eq. (\ref{93}) and take ${\cal R}$ as lower
bound in the second term. Computing each of these integrals we find

\begin{equation}\label{94}
\sum\frac{{\hat x}_L}{|{\mathbf
x}|^{2l+1}}\left(\frac{d}{dB}\right)^p\left[\frac{|{\mathbf
x}|^{B+a+l+3}-{\cal R}^{B+a+l+3}}{B+a+l+3}\right] + \sum {\hat
x}_L\left(\frac{d}{dB}\right)^p\left[\frac{-|{\mathbf
x}|^{B+a-l+2}}{B+a-l+2}\right]
\;. \end{equation}
Each of these terms clearly admits an analytic continuation for any
$B\in{\cal B}_\epsilon$ and in fact for any $B\in\C$ except at
integer values. Furthermore we see from that expression that the
function will admit a Laurent expansion when $B\to 0$, with in general
some multiple poles [coming from the differentiation $(d/dB)^p$ of
simple poles $\sim 1/B$]. Hence our statement.

It is clear that the Laplacians of the two integrals $I_<^B$ and
$I_>^B$ satisfy, in the domains of the complex plane where these
functions were initially valid~:

\begin{eqnarray}\label{95}
\Re(B)>-3~~~\Longrightarrow~~~\Delta I_<^B({\mathbf x},t) &=& Y({\cal
R}-|{\mathbf x}|)~\!|\widetilde{{\mathbf x}}|^B~\!\pn{\tau}({\mathbf x},t)\;,\\
\Re(B)<-a_\mathrm{max}-2~~~\Longrightarrow~~~\Delta I_>^B({\mathbf x},t) &=&
Y(|{\mathbf x}|-{\cal R})~\!|\widetilde{{\mathbf x}}|^B~\!\pn{\tau}({\mathbf x},t)
\;, \end{eqnarray}
where $Y$ denotes the Heaviside step-function. Therefore, if we {\it
define} for any $B\in{\cal B}_\epsilon$ the object

\begin{equation}\label{96}
I^B({\mathbf x},t) = I_<^B({\mathbf x},t) + \un{B\in{\cal
B}_\epsilon}{\hbox{analytic continuation}}~\!  \Big\{I_>^B({\mathbf
x},t)\Big\}
\;, \end{equation}
we find that it necessarily satisfies, for any $B\in{\cal
B}_\epsilon$, the $B$-dependent Poisson equation

\begin{equation}\label{97}
\Delta I^B({\mathbf x},t) = |\widetilde{{\mathbf x}}|^B~\!\pn{\tau}({\mathbf x},t)
\;. \end{equation}
On the other hand, we have learned from Eq. (\ref{94}) that $I^B$
admits when $B\to 0$ a Laurent expansion involving (in general) simple
and multiple poles. Now the key idea, as we shall prove, is that the
{\it finite part}, or coefficient of the zero-th power of $B$ in the
latter Laurent expansion, represents a particular solution of the
Poisson equation that we want to solve.  Let the Laurent expansion of
$I^B$ be

\begin{equation}\label{98}
I^B({\mathbf x},t) = \sum_{k=k_\mathrm{min}}^{+\infty} i_k({\mathbf x},t)B^k
\;, \end{equation}
where $k_\mathrm{min}\in \Z$, and where the coefficients $i_k$
depend on the field point $({\mathbf x},t)$. By applying the Laplacian
operator onto both sides of Eq. (\ref{98}), and using the result
(\ref{97}) together with the Taylor expansion of the regularization
factor $|\widetilde{{\mathbf x}}|^B$, we arrive at

\begin{eqnarray}\label{99}
k_\mathrm{min}\leq k\leq -1~~\Longrightarrow~~\Delta i_k &=& 0 \;,\\ k\geq
0~~\Longrightarrow~~\Delta i_k &=& \frac{(\ln |\widetilde{{\mathbf
x}}|)^k}{k!}~\!\pn{\tau}
\;. \end{eqnarray}
Thus, the case $k=0$ shows that the finite-part coefficient in the
expansion (\ref{98}), namely $i_0$, is a particular solution of the
required equation~: $\Delta i_0=\pn{\tau}$. We shall now forget about
the intermediate name $i_0$, and denote, from now on, the latter
solution by $\Dt\pn{\tau}\equiv i_0$, or, in more explicit terms,

\begin{equation}\label{100}
\Dt \pn{\tau}({\mathbf x},t) = \pf\Delta^{-1}\Big[|\widetilde{{\mathbf
x}}|^B\pn{\tau}({\mathbf x},t)\Big]
\;, \end{equation} 
where $\Delta^{-1}$ refers to the standard Poisson integral, and the
finite-part symbol $\pf$ means the previous operations of considering
the Laurent expansion when $B\to 0$, and picking up the finite-part
coefficient. Thus, we have proved that $\Delta\big[\Dt \pn{\tau}\big]
= \pn{\tau}$, so the generalized inverse Poisson operator $\Dt$
defines a particular solution of the Poisson equation, which has, by
construction, none of the problems of divergencies of Poisson
integrals which have so much plagued the standard post-Newtonian
approximation
\cite{C65,CN69,CE70,AD75,Ehl77,Ehl80,Ker80,Ker80',Capo81,PapaL81,BRu81,BRu82}.

Finally let us prove that our generalized solution $\Dt \pn{\tau}$
owns the same properties (\ref{87})-(\ref{88}) as the corresponding
source $\pn{\tau}$. This verification is important because it will
allow us to iterate any number of times the operator $\Dt$, and to
obtain the post-Newtonian expansion up to any post-Newtonian order.
The main problem amounts to prove that $\Dt \pn{\tau}$ admits the same
type of expansion at infinity $|{\mathbf x}|\to +\infty$ as in
Eq. (\ref{88}). To do this we consider again the same split into
near-zone and far-zone contributions~: $\Dt \pn{\tau} = I_< + I_>$,
where

\begin{eqnarray}\label{102}
I_<({\mathbf x},t) &=& -\frac{1}{4\pi}\pf\int_{|{\mathbf y}|<{\cal
R}}\frac{d^3{\mathbf y}}{|{\mathbf x}-{\mathbf y}|} ~\!|\widetilde{{\mathbf
y}}|^B\pn{\tau}({\mathbf y},t)\label{102a}\;,\\ I_>({\mathbf x},t) &=&
-\frac{1}{4\pi}\pf\int_{|{\mathbf y}|>{\cal R}}\frac{d^3{\mathbf y}}{|{\mathbf
x}-{\mathbf y}|} ~\!|\widetilde{{\mathbf y}}|^B\pn{\tau}({\mathbf
y},t)\label{102b}
\;. \end{eqnarray}
The near-zone integral admits an expansion at infinity which is of the
required type.  Indeed, because the integrand is of compact support,
$|{\mathbf y}|<{\cal R}$, we can replace in it the factor $\frac{1}{|{\mathbf
x}-{\mathbf y}|}$ by its expansion
$\sum\frac{(-)^l}{l!}y^L\hat{\partial}_L\frac{1}{|{\mathbf
x}|}$ and integrate term by term. So we have, $\forall N\in\N$,

\begin{eqnarray}\label{103}
I_<({\mathbf x},t) &=&
-\frac{1}{4\pi}\sum_{l=0}^{N-1}\frac{(-)^l}{l!}\hat{\partial}_L\left(\frac{1}{r}\right)
\pf\int_{|{\mathbf y}|<{\cal R}}d^3{\mathbf y}~\!|\widetilde{{\mathbf
y}}|^By^L\pn{\tau}({\mathbf y},t)+{\cal O}\left(\frac{1}{r^N}\right)
\;. \end{eqnarray}
The right-hand-side has indeed the same structure as in
Eq. (\ref{88}).  The treatment of the far-zone integral is more
delicate. We proceed in a way similar to what was done in
Eqs. (\ref{90})-(\ref{94}). Namely we replace into it the source
$\pn{\tau}$ by its expansion given by Eq. (\ref{88}). This yields (the
finite part of) Eq. (\ref{90}), that for convenience we reproduce
here~:

\begin{equation}\label{104}
I_>({\mathbf x},t) = -\frac{1}{4\pi}\pf\int_{|{\mathbf y}|>{\cal
R}}\frac{d^3{\mathbf y}~\!|\widetilde{{\mathbf y}}|^B}{|{\mathbf x}-{\mathbf
y}|}\left\{\sum {\hat n}_L({\mathbf y}) |{\mathbf y}|^a (\ln |{\mathbf y}|)^p
G_{L,a,p}(t) + S_N({\mathbf y},t)\right\}
\;. \end{equation}
There is a contribution of the remainder and a finite sum of terms
with known structure.  The remainder contribution is simply given by
the value at $B=0$ which has been written in the right-hand-side of
Eq. (\ref{91}). Let us write this term in the form

\begin{equation}\label{105}
\int_{|{\mathbf y}|>{\cal R}}\frac{d^3{\mathbf y}}{|{\mathbf x}-{\mathbf y}|} S_N({\mathbf
y},t) =
\sum_{l=0}^{N-4}\frac{(-)^l}{l!}\hat{\partial}_L\left(\frac{1}{r}\right)\int_{|{\mathbf
y}|>{\cal R}}d^3{\mathbf y}~\!y^L S_N({\mathbf y},t)+ T_{N-2}({\mathbf x},t)
\;, \end{equation}
where we introduced the $N-4$ first terms of the multipolar expansion
of $\frac{1}{|{\mathbf x}-{\mathbf y}|}$ when $r=|{\mathbf x}|\to +\infty$, and
where

\begin{equation}\label{106}
T_{N-2}({\mathbf x},t) = \int_{|{\mathbf y}|>{\cal R}}d^3{\mathbf
y}\left[\frac{1}{|{\mathbf x}-{\mathbf y}|}-\sum_{l=0}^{N-4}
\frac{(-)^l}{l!}y^L\hat{\partial}_L\left(\frac{1}{r}\right)\right]S_N({\mathbf
y},t)
\;. \end{equation}
The maximal order $N-4$ of the expansion is chosen in such a way that
all the terms in Eq. (\ref{106}) are given by convergent integrals at
infinity, owing to the fact that the remainder satisfies $S_N={\cal
O}(1/r^N)$. Now we prove that $T_{N-2}$, defined by Eq. (\ref{106}),
is also a remainder in the sense that $T_{N-2}={\cal O}(\ln
r/r^{N-2})$.  We split $T_{N-2}$ into two integrals, a near-zone
integral $T^\mathrm{near}_{N-2}$ corresponding to the integration range
$|{\mathbf y}|\in \big]{\cal R}, |{\mathbf x}|\big[$, and a far-zone one
$T^\mathrm{far}_{N-2}$ corresponding to $|{\mathbf y}|\in \big]|{\mathbf x}|,
+\infty\big[$.  In the near-zone integral we can use the bound

\begin{equation}\label{107}
\left|\frac{1}{|{\mathbf x}-{\mathbf y}|}-\sum_{l=0}^{N-4}
\frac{(-)^l}{l!}y^L\hat{\partial}_L\left(\frac{1}{r}\right)\right|\leq
C_N\frac{|{\mathbf y}|^{N-3}}{|{\mathbf x}|^{N-2}}
\;, \end{equation}
where $C_N$ is a constant. On the other hand, because
$S_N={\cal O}(1/r^N)$, there is also a constant $A_N$, depending on
the value of ${\cal R}$, such that the following majoration holds~:

\begin{equation}\label{108}
\left|S_N({\mathbf y},t)\right|\leq \frac{A_N}{|{\mathbf y}|^N}
\;. \end{equation}
Replacing these results into the near-zone integral we get

\begin{equation}\label{109}
\left|T^\mathrm{near}_{N-2}({\mathbf x},t)\right|\leq 4\pi
~\!\frac{A_NC_N}{|{\mathbf x}|^{N-2}}~\!\ln\left(\frac{|{\mathbf x}|}{{\cal
R}}\right)
\;. \end{equation}
In the far-zone integral, we can no longer apply the bound (\ref{107})
but still we can employ the majoration (\ref{108}). Then we can easily
show the inequality (in which $|{\mathbf y}|=|{\mathbf x}|\lambda$)

\begin{equation}\label{110}
\left|T^\mathrm{far}_{N-2}({\mathbf x},t)\right|\leq 4\pi~\!\frac{A_N}{|{\mathbf
x}|^{N-2}}~\!\int_1^{+\infty}\frac{d\lambda}{\lambda^{N-2}}\left[\frac{1}{\lambda}
+\sum_{l=0}^{N-4}\frac{(2l-1)!!}{l!}\lambda^l\right]
\;. \end{equation}
The integral is convergent. At last, from Eqs. (\ref{109}) and
(\ref{110}) we have proved that $T_{N-2}={\cal O}(\ln r/r^{N-2})$.
Still it remains to show that the finite sum of terms in
Eq. (\ref{104}), i.e. besides the remainder, admits some expansions of
the required structure. But this follows from applying the finite part
operation $\pf$ onto the result (\ref{94}), which tells us immediately
that we have an expansion of the correct type $\sim{\hat n}_L({\mathbf
x})|{\mathbf x}|^a(\ln |{\mathbf x}|)^q$.

\section{Far zone expansion of the Poisson integral}\label{C}

Thanks to the investigation in Appendix \ref{B}, the far-zone (or
multipolar) expansion of the object $\Dt[\pn{\tau}]$ happens to be
workable. Recall that controlling the far-zone expansion of the
post-Newtonian field is fundamental since it is at the basis of the
matching. The operation of taking the far-zone expansion is denoted
${\cal M}$ when applied on post-Newtonian objects (see Section
\ref{IIIB}). We therefore want to determine the expression of ${\cal
M}(\Dt [\pn{\tau}])$. That is, we want to relate it to the expansion
of the corresponding source, which has the same structure as in
Eq. (\ref{51})~:

\begin{equation}\label{111}
{\cal M}(\pn\tau)({\mathbf x},t) =~ \sum {\hat n}_L r^a (\ln r)^p
G_{L,a,p}(t)
\;. \end{equation}
By the matching equation we know that this far-zone expansion is
identical with the near-zone expansion of the external field [see
e.g. Eq. (\ref{69})].  Let us first apply ${\cal M}$ onto
$\Dt[\pn{\tau}]$ as expressed as a sum of near-zone and far-zone
contributions,

\begin{equation}\label{112}
{\cal M}\left(\Dt [\pn{\tau}]\right) ={\cal M}\big(I_<\big)
+{\cal M}\big(I_>\big)
\;, \end{equation}
where $I_<$ and $I_>$ are defined by
Eqs. (\ref{102a})-(\ref{102b}). The near-zone integral is quite easy
to work with.  Indeed, from Eq. (\ref{103}) we see that its expansion
when $r=|{\mathbf x}|\to +\infty$ is obtained by expanding the factor
$1/|{\mathbf x}-{\mathbf y}|$ inside the integrand.  Therefore, the infinite
far-zone expansion (without remainder) reads

\begin{equation}\label{113}
{\cal M}\big(I_<\big) = -\frac{1}{4\pi} \pf\int_{|{\mathbf y}|<{\cal
R}}d^3{\mathbf y}~\!|\widetilde{{\mathbf y}}|^B{\cal M}\left(\frac{1}{|{\mathbf
x}-{\mathbf y}|}\right)\pn{\tau}({\mathbf y},t)
\;, \end{equation}
in which we denote 

\begin{equation}\label{114}
{\cal M}\left(\frac{1}{|{\mathbf x}-{\mathbf y}|}\right) =
\sum_{l=0}^{+\infty}\frac{(-)^l}{l!} y_L
\hat{\partial}_L\!\left(\frac{1}{|{\mathbf x}|}\right)
\;. \end{equation}
On the other hand, the far-zone expansion of the far-zone integral
$I_>$ has been obtained in Eq. (\ref{104})-(\ref{105}), where we found
that it comes from replacing the source term by its far-zone
expansion [indeed, when ${\cal R}$ is large enough, the integration
ranges over the domain of validity of the far-zone expansion]. So the
infinite far-zone expansion of that term is given by
 
\begin{equation}\label{115}
{\cal M}\big(I_>\big) = -\frac{1}{4\pi} \pf\int_{|{\mathbf y}|>{\cal R}}
\frac{d^3{\mathbf y}~\!|\widetilde{{\mathbf y}}|^B}{|{\mathbf x}-{\mathbf y}|}~\!{\cal
M}\Big(\pn{\tau}({\mathbf y},t)\Big)
\;, \end{equation}
where the integrand contains the expansion of the source given by
Eq. (\ref{111}). Now let us use a technical lemma which is quite
important in the present formalism, and has already played a crucial
role in Refs. \cite{B95,B98mult}. This lemma is based on the remark
that any radial integral of the type $\int_0^{+\infty}d|{\mathbf y}||{\mathbf
y}|^{B+a}(\ln |{\mathbf y}|)^p$, where $B\in\C$ and $a$ and $p$ are
arbitrary real numbers, is identically {\it zero} by analytic
continuation in $B$. See Ref. \cite{N6} for the proof. Our useful
lemma, that is trivial to relate to the previous remark (after
performing the integration over angles), is

\begin{equation}\label{116}
\pf\int d^3{\mathbf y}~\!|\widetilde{{\mathbf y}}|^B{\cal
M}\left(\frac{1}{|{\mathbf x}-{\mathbf y}|}\right){\cal
M}\Big(\pn{\tau}({\mathbf y},t)\Big) =0 \;. \end{equation} The point
here is that the integral ranges over the complete three-dimensional
space $\Real^3$. Now we have the ``numerical'' equalities ${\cal
M}\big(\frac{1}{|{\mathbf x}-{\mathbf y}|}\big)=\frac{1}{|{\mathbf
x}-{\mathbf y}|}$ when $|{\mathbf y}|<|{\mathbf x}|$ and ${\cal
M}\left(\pn{\tau}\right)=\pn{\tau}$ when $|{\mathbf y}|>a$, where $a$
is the radius of the compact-support source. From this we deduce that
as soon as ${\cal R}>a$, what we can always assume right from the
beginning, and $|{\mathbf x}|>{\cal R}$, which is not a problem
because we are considering the limit $|{\mathbf x}|\to +\infty$, we
have the identity

\begin{equation}\label{117}
\pf\int_{|{\mathbf y}|<{\cal R}} \frac{d^3{\mathbf y}~\!|\widetilde{{\mathbf
y}}|^B}{|{\mathbf x}-{\mathbf y}|}~\!{\cal M}\Big(\pn{\tau}({\mathbf y},t)\Big) +
\pf\int_{|{\mathbf y}|>{\cal R}}d^3{\mathbf y}~\!|\widetilde{{\mathbf y}}|^B{\cal
M}\left(\frac{1}{|{\mathbf x}-{\mathbf y}|}\right)\pn{\tau}({\mathbf y},t) =0
\;. \end{equation}
By means of that identity we can obtain the requested form of the
far-zone expansion as

\begin{equation}\label{118}
{\cal M}\left(\Dt [\pn{\tau}]\right) = -\frac{1}{4\pi}\pf\int
d^3{\mathbf y}~\!|\widetilde{{\mathbf y}}|^B\left[\frac{1}{|{\mathbf
x}-{\mathbf y}|}{\cal M}\Big(\pn{\tau}({\mathbf y},t)\Big) +{\cal
M}\left(\frac{1}{|{\mathbf x}-{\mathbf y}|}\right)\pn{\tau}({\mathbf
y},t)\right] \;. \end{equation} In this particular form we see that
the ${\cal M}$-operator is distributed on the two terms like a
derivative operator would be. In the first term we recognize the
action of the generalized Poisson integral.  Actually this Poisson
operator has been defined in Appendix \ref{A} when acting on a
near-zone expansion of the type (\ref{69}), but by matching that
expansion is the same as the present far-zone expansion, so the
definition is rigorously the same.  Finally we can re-write
Eq. (\ref{118}) into the alternative form

\begin{equation}\label{119}
{\cal M}\left(\Dt [\pn{\tau}]\right) = \Dt\left[{\cal
M}(\pn{\tau})\right] -
\frac{1}{4\pi}\sum_{l=0}^{+\infty}\frac{(-)^l}{l!}\hat{\partial}_L\!\left(r^{-1}\right)
\pf\int d^3{\mathbf y}~\!|\widetilde{{\mathbf y}}|^B\hat{y}_L\pn{\tau}({\mathbf
y},t)
\;, \end{equation}
which constitutes the main result of this Appendix. Notice that
Eq. (\ref{119}) is in agreement with the multipole expansion of the
retarded integral as given by Eq. (3.11)-(3.12) in
Ref. \cite{B98mult}, when specialized to the static case where there
is no dependence on time.
 
Next we derive the analogous result concerning the operator of the
``instantaneous'' potentials

\begin{equation}\label{120}
\widetilde{\im} =
\sum_{k=0}^{+\infty}\frac{1}{c^{2k}}\partial_t^{2k}\Dn{-k-1}
\;. \end{equation}
We iterate $k+1$ times the result (\ref{119}). There is no problem for
doing this; the only point is that we use in a repeated way the easily
checked formula telling that we are allowed to ``operate by parts''
the Poisson integral $\Dt$ in the way

\begin{equation}
\pf\int d^3{\mathbf z}~\!|\widetilde{{\mathbf
z}}|^B\hat{z}_L\Dt\pn{\tau}=\pf\int d^3{\mathbf y}~\!|\widetilde{{\mathbf
y}}|^B\Dt\left[\hat{y}_L\right]\pn{\tau}
\;. \end{equation}
This formula is a consequence of the fact that $\pf\int d^3{\mathbf
z}~\!|\widetilde{{\mathbf z}}|^B\frac{\hat{z}_L}{|{\mathbf z}-{\mathbf y}|}
=-4\pi\Dt\left[\hat{y}_L\right]=\frac{-2\pi}{2l+3}|{\mathbf y}|^2\hat{y}_L$; see Eq. (4.10) in
Ref. \cite{B96}.  Therefore we arrive at

\begin{eqnarray}\label{121}
{\cal M}\left(\Dn{-k-1} ~\![\pn{\tau}]\right) &=& \Dn{-k-1}\left[{\cal
 M}(\pn{\tau})\right]\\ &-&
 \frac{1}{4\pi}\sum_{l=0}^{+\infty}\frac{(-)^l}{l!}
 \sum_{i=0}^{k}\Dn{-i}\left[\hat{\partial}_L\!\left(r^{-1}\right)\right]
 \pf\int d^3{\mathbf y}~\!|\widetilde{{\mathbf
 y}}|^B\Dn{i-k}\left[\hat{y}_L\right]\pn{\tau}({\mathbf y},t)\nonumber
\;, \end{eqnarray}
and from this it is very simple to derive the requested expression
concerning $\widetilde{\im}$. We obtain

\begin{eqnarray}\label{122}
{\cal M}\left(\widetilde{\im}\left[\pn{\tau}\right]\right) &=&
 \widetilde{\im}[{\cal M}(\pn{\tau})] \\ &-&
 \frac{1}{4\pi}\sum_{l=0}^{+\infty}\frac{(-)^l}{l!}
 \sum_{i=0}^{+\infty}\Dn{-i}\left[\hat{\partial}_L\!
\left(r^{-1}\right)\right]\sum_{k=i}^{+\infty}\frac{1}{c^{2k}}
 \pf\int d^3{\mathbf y}~\!|\widetilde{{\mathbf
 y}}|^B\Dn{i-k}\left[\hat{y}_L\right]\partial_t^{2k}\pn{\tau}({\mathbf
 y},t)\nonumber
\;. \end{eqnarray} 
This expression, though completely explicit, does not constitute our
final form. Because the ``instantaneous'' solution is a particular
solution of the d'Alembertian equation, it must be possible to
re-express the second term in Eq. (\ref{122}) as a combination of some
source-free retarded and advanced multipolar waves. To see this we
notice that

\begin{equation}\label{123}
\Dn{-i}\left[\hat{\partial}_L\!\left(r^{-1}\right)\right]
=\hat{\partial}_L\left(\frac{r^{2i-1}}{(2i)!}\right)
\;, \end{equation}
which shows that the latter homogeneous solution is actually one of
the {\it symmetric} type, i.e. retarded {\it plus} advanced. Namely we
can re-write Eq. (\ref{122}) into the form

\begin{equation}\label{124}
{\cal M}\left(\widetilde{\im}\left[\pn{\tau}\right]\right) =
 \widetilde{\im}[{\cal M}(\pn{\tau})] - \frac{1}{4\pi}
 \sum^{+\infty}_{l=0} {(-)^l\over l!} \hat{\partial}_L \left\{
 \frac{\overline{{\cal F}_L (t-r/c)+{\cal F}_L (t+r/c)}}{2r} \right\}
 \;, \end{equation} where the overline notation means taking the
 Taylor expansion of the symmetric wave when the retardation $r/c\to
 0$ [the result is displayed in Eq. (\ref{53})]. Actually, this
 overline notation is somewhat misleading, because, in keeping with
 the real meaning of the result (\ref{124}), one should {\it a
 posteriori} interpret the latter Taylor expansion as a {\it far-zone}
 (singular) expansion when $r\to +\infty$. However, in view of the
 matching, it is more fruitful to employ the same overline notation as
 for the expansion of the anti-symmetric waves occuring in the
 near-zone metric -- indeed the matching is simply interested at
 identifying together some asymptotic expansions which are of the same
 form. The ``multipole-moment'' function ${\cal F}_L (t)$ in
 Eq. (\ref{124}) is given by

\begin{equation}\label{125}
{\cal F}_L (t) = \sum_{j=0}^{+\infty}\frac{1}{c^{2j}} \pf\int d^3{\mathbf
y}~\!|\widetilde{{\mathbf
y}}|^B\Dn{-j}\left[\hat{y}_L\right]\partial_t^{2j}\pn{\tau}({\mathbf y},t)
\;. \end{equation}
Finally let us find an alternative, more compact, form for this
result. We introduce the $l$-dependent function

\begin{equation}\label{126}
\delta_l (z) = {(2l+1)!!\over 2^{l+1} l!} (1-z^2)^l
\;, \end{equation} 
whose integral is normalized to one~: $\int^1_{-1} dz~\!\delta_l (z) =
1$. One can readily show that

\begin{equation}\label{127}
\Dn{-j}\left[\hat{y}_L\right] = |{\mathbf y}|^{2j}{\hat y}_L
\int_{-1}^1dz~\!\frac{z^{2j}}{(2j)!}\delta_l(z)
\;, \end{equation}
which permits to express the function ${\cal F}_L$ in a form where the
post-Newtonian series is formally re-summed as

\begin{equation}\label{128}
{\cal F}_L (t) = \pf\int d^3 {\mathbf x}~\! |\widetilde{\mathbf y}|^B
 {\hat y}_L \, \overline{\int^1_{-1} dz ~\!\delta_l(z)~\!\pn{\tau}
 ({\mathbf y}, t\pm z|{\mathbf y}|/c)} \;. \end{equation} Under this form
 we recognize the multipole-moment function introduced in Eq. (3.14)
 in Ref. \cite{B98mult} (the function remains unchanged by taking
 either sign $\pm$ in the time argument of $\pn{\tau}$). This result
 permits to fully determine the exterior multipolar field by matching,
 and to recover the expression already obtained in Ref. \cite{B98mult}
 by means of a somewhat different method.


\begin{references}
\bibitem{N4}Our notation for STF tensors is the
following. $L=i_1i_2\cdots i_l$ denotes a multi-index, made of $l$
(spatial) indices. When summing over multi-indices we never write the
$l$ summations over the $l$ indices $i_1,\cdots ,i_l$ ranging from 1
to 3. The STF product of unit vectors $n_i=n^i\equiv x^i/r$ is denoted
${\hat n}_L$=STF($n_L$), where $n_L$ is a short-hand for
$n_{i_1}\cdots n_{i_l}$. For instance ${\hat
n}_{ij}=n_in_j-\frac{1}{3}\delta_{ij}$. Similarly we denote
$x_L=x_{i_1}\cdots x_{i_l}=r^l n_{L}$ and ${\hat x}_L$=STF($x_L$). The
derivative operator $\partial_L$ is a short-hand for
$\partial_{i_1}\cdots\partial_{i_l}$, and we 
have ${\hat \partial}_L$=STF($\partial_L$). For instance ${\hat
\partial}_{ij}=\partial_{ij}-\frac{1}{3}\delta_{ij}\Delta$.  More
generally, a function $F_L$ is said to be STF with respect to the
$l$ indices composing $L$ if and only if, for any pair of indices
$i_p,i_q\in L$, we have $F_{\cdots i_p\cdots i_q\cdots}=F_{\cdots
i_q\cdots i_p\cdots}$ and $\delta_{i_pi_q}F_{\cdots i_p\cdots
i_q\cdots}=0$ (see the Appendices A and B in Ref. \cite{BD86} for
reviews about the STF formalism).
\bibitem{NN}It is clear that for stationary sources (independent of
time), the anti-symmetric waves given by Eqs. (\ref{24}) are
zero. Therefore the only contribution to the function ${\cal
R}\mn_L(t)$ comes from the non-stationary (or radiative) part of the
field, which according to our assumption of stationarity in the past
is zero when $t\leq -{\cal T}$, and for which ${\cal R}\mn_L(t)$ is perfectly
well-defined. For simplicity in the notation we do not indicate that
${\cal R}\mn_L(t)$ should be computed only from the ``radiative'' part
of the source term ${\cal M}(\tau\mn)$.
\bibitem{N6} We want to prove that the radial integral
$\int_0^{+\infty} d\absy~\!\absy^{B+a} (\ln \absy)^p$ is zero by
analytic continuation ($\forall B\in\C$).  First we can get rid
of the logarithms by considering some repeated differentiations with
respect to $B$; thus we need only to consider the simpler integral
$\int_0^{+\infty} d\absy~\!\absy^{B+a}$. We split the integral into a
near-zone integral $\int_0^{\cal R} d\absy~\!\absy^{B+a}$ and a
far-zone one $\int_{\cal R}^{+\infty} d\absy~\!\absy^{B+a}$, where
${\cal R}$ is some constant radius. When $\Re (B)$ is a large enough
positive number, the value of the near-zone integral is ${\cal
R}^{B+a+1}/(B+a+1)$, while when $\Re (B)$ is a large negative number,
the far-zone integral reads the opposite, $-{\cal R}^{B+a+1}/(B+a+1)$.
Both obtained values represent the unique analytic continuations of
the near-zone and far-zone integrals for any $B\in\C$ except
$-a-1$. The complete integral $\int_0^{+\infty} d\absy~\!\absy^{B+b}$
is equal to the sum of these analytic continuations, and is therefore
identically zero ($\forall B\in\C$, including the value $-a-1$).
\end{references}
\end{document}